\documentclass[a4paper, twoside, 12pt]{article}

\usepackage[noblocks]{authblk} 
\usepackage[numbers]{natbib}
\usepackage{graphicx}
\usepackage{fancyhdr}
\usepackage[colorlinks=true, linkcolor=MidnightBlue, urlcolor=MidnightBlue, citecolor=PineGreen]{hyperref}
\usepackage[tableposition=top, justification=justified, textfont=it]{caption}
\usepackage{titlesec}
\usepackage{multicol}
\usepackage{afterpage}
\usepackage{lipsum}
\usepackage[dvipsnames, x11names]{xcolor}
\usepackage[inner=4cm, outer=4cm, top=2.8cm, bottom=2.8cm, marginparsep=0cm, marginparwidth=0cm, includehead, includefoot]{geometry}
\usepackage{setspace}
\usepackage{array}
\usepackage{bm}
\usepackage{tikz}
\usepackage{float}
\usepackage[framemethod]{mdframed}
\usepackage{amsmath}
\usepackage{amssymb}

\titleformat{\section}{\Large \bfseries \centering \scshape}{\thesection.}{0.3em}{}[{\titlerule[0.5pt]}]

\definecolor{shadecolor}{RGB}{230,230,230}
\newcommand{\mybox}[1]{\par\noindent\colorbox{shadecolor}
{\parbox{\dimexpr\textwidth-2\fboxsep\relax}{#1}}}

\titleformat{\subsection}{\large \bfseries \mybox}{\thesubsection}{1em}{}

\titleformat{\subsubsection}{\itshape}{\thesubsubsection.}{0.3em}{}

\renewenvironment{abstract}
{\vskip 2.5ex {\large\bf\noindent Abstract}\vspace{0.7ex} \\ %
  \bgroup\noindent\ignorespaces}%
{\par\egroup\vskip 2.5ex}

\newenvironment{keywords}
{\bgroup\leftskip 20pt\rightskip 20pt \small\noindent{\bf Keywords:} }%
{\par\egroup\vskip 10ex}

\fancypagestyle{firstpage}{%
  
  \fancyhead[]{}
  \fancyfoot[C]{}
  \fancyfoot[L]{\footnotesize To appear in \\
  O. Colliot (Ed.), \textit{Machine Learning for Brain Disorders}, Springer\\
  }

}

\fancypagestyle{otherpages}{%
  
  \fancyhead[LE]{\thepage}
  \fancyhead[RE]{\runningauthor}
  \fancyhead[LO]{\runningheadtitle}
  \fancyhead[RO]{\thepage}
  \fancyfoot[L]{\footnotesize  \textit{Machine Learning for Brain Disorders}, Chapter~\chapternumber}
  \fancyfoot[C]{}
}

\makeatletter
\renewcommand{\maketitle}{\bgroup\setlength{\parindent}{0pt}

\begin{flushright}
  \color{MidnightBlue}
  \textbf{\LARGE Chapter~\chapternumber}
\end{flushright}

\vspace{0.3in}

\begin{flushleft}
    \setstretch{2.0} 
    \textbf{\color{MidnightBlue}\huge\@title}
\end{flushleft}

\vspace{0.15in}

\begin{flushleft}
    \textbf{\bfseries \large\@author}
\end{flushleft}\egroup
}

\makeatother

\DeclareCaptionLabelFormat{labelformat}{\color{MidnightBlue}\bfseries #1 #2}
\renewcommand{\bibpreamble}{\scriptsize \begin{multicols}{2}}
\renewcommand{\bibpostamble}{\end{multicols}}

\mdfsetup{skipabove=\topskip, skipbelow=\topskip}

\newcounter{nicebox}

\newfloat{floatbox}{thp}{lop}
\floatname{floatbox}{Box}

\DeclareMathAlphabet{\mathsfit}{\encodingdefault}{\sfdefault}{m}{sl}
\SetMathAlphabet{\mathsfit}{bold}{\encodingdefault}{\sfdefault}{bx}{n}

\usepackage{xurl}
\usepackage[utf8]{inputenc}

\begin{document}


\newcommand{\runningauthor}{Jimenez \textit{and} Racoceanu} 

\newcommand{\runningheadtitle}{Computational Pathology for Brain Disorders}

\newcommand{\chapternumber}{18}

\newcommand{\emailaddress}{daniel.racoceanu@sorbonne-universite.fr}

\title{Computational Pathology for Brain Disorders} 

\author[1]{Gabriel Jiménez}
\author[*,1]{Daniel Racoceanu}  

\affil[1]{Sorbonne Université, Institut du Cerveau - Paris Brain Institute - ICM, CNRS, Inria, Inserm, AP-HP, Hôpital Pitié Salpêtrière, 75013, Paris, France}

\affil[*]{Corresponding author: e-mail address: \href{mailto:\emailaddress}{\emailaddress}}

\maketitle

\afterpage{\aftergroup\restoregeometry}
\pagestyle{otherpages}

\begin{abstract}
Non-invasive brain imaging techniques allow understanding the behavior and macro changes in the brain to determine the progress of a disease. However, computational pathology provides a deeper understanding of brain disorders at cellular level, able to consolidate a diagnosis and make the bridge between the medical image and the omics analysis.
In traditional histopathology, histology slides are visually inspected, under the microscope, by trained pathologists. This process is time-consuming and labor-intensive; therefore, the emergence of Computational Pathology has triggered great hope to ease this tedious task and make it more robust. 
This chapter focuses on understanding the state-of-the-art machine learning techniques used to analyze whole slide images within the context of brain disorders. We present a selective set of remarkable machine learning algorithms providing discriminative approaches and quality results on brain disorders. These methodologies are applied to different tasks, such as monitoring mechanisms contributing to disease progression and patient survival rates, analyzing morphological phenotypes for classification and quantitative assessment of disease, improving clinical care, diagnosing tumor specimens, and intraoperative interpretation. 
Thanks to the recent progress in machine learning algorithms for high-content image processing, computational pathology marks the rise of a new generation of medical discoveries and clinical protocols, including in brain disorders.
\end{abstract}


\section{Introduction}
\label{sec:intro}

\subsection{What are we presenting?}

This chapter aims to assist the reader in discovering and understanding state-of-the-art machine learning techniques used to analyze Whole Slide Images (WSI), an essential data type used in Computational Pathology (CP). 
We are restricting our review to brain disorders, classified within four generally-accepted groups:
\begin{itemize}
    \item \textit{Brain injuries:} caused by blunt trauma and can damage brain tissue, neurons, and nerves.
    \item \textit{Brain tumors:} can originate directly from the brain (and be cancerous or benign) or be due to metastasis (cancer elsewhere in the body and spreading to the brain).
    \item \textit{Neurodegenerative diseases:} brain and nerves deteriorate over time. We include, here, Alzheimer's disease, Huntington's disease, ALS (amyotrophic lateral sclerosis) or Lou Gehrig's disease, Parkinson's disease.
    \item \textit{Mental disorders:} (or mental illness) affect behavior patterns. Depression, anxiety, bipolar disorder, PTSD (Post-traumatic stress disorder), and schizophrenia are common diagnoses.
\end{itemize}

In the last decade, there has been exponential growth in the application of image processing and artificial intelligence (AI) algorithms within digital pathology workflows. First FDA (U.S. Food and Drug Administration) clearance of digital pathology for diagnosis protocols was as early as 2017 \footnote{\href {https://www.fda.gov/news-events/press-announcements/fda-allows-marketing-first-whole-slide-imaging-system-digital-pathology}{https://www.fda.gov/news-events/press-announcements}}, as the emergence of innovative deep learning (DL) technologies have made this possible, with the requested degree of robustness and repeatability. 

Ahmed Serag \textit{et al.} \cite{Serag2019-ug} discuss the translation of AI into clinical practice to provide pathologists with new tools to improve diagnostic consistency and reduce errors. In the last five years, the authors reported an increase in academic publications (over 1000 articles reported in PubMed) and over \$100M invested in start-ups building practical AI applications for diagnostics. The three main areas of development are: (i) \textit{network architectures} to extract relevant features from WSI for classification or segmentation purposes; (ii) \textit{generative adversarial networks (GANs)} to address some of the issues present in the preparation and acquisition of WSIs; and, (iii) \textit{unsupervised learning} to create labeling tools for precise annotations. Regarding data, many top-tier conference competitions have been organized and released annotated datasets to the community; however, very few of them contain brain tissue samples. Those which do, are from brain tumor regions obtained during a biopsy, making it harder to study other brain disorder categories which frequently require post-mortem data. 

In \cite{Serag2019-ug}, the authors also mention seven key challenges in diagnostic AI in pathology, listed as follows:

\begin{itemize}
    \item Access to large well-annotated data sets. Most articles on brain disorders use private datasets due to hospital privacy constraints.
    \item Context switching between workflows refers to a seamless integration of AI into the pathology workflow.
    \item Algorithms are slow to run as image sizes are in gigapixels' order and require considerable computational memory.
    \item Algorithms require configuration, and fully automated approaches with high accuracy are difficult to develop.
    \item Properly defined protocols are needed for training and evaluation.
    \item Algorithms are not properly validated due to a lack of open datasets. However, research in data augmentation might help in this regard.
    \item Introduction of Intelligence Augmentation to describe computational pathology improvements in diagnostic pathology. AI algorithms work best on well-defined domains rather than in the context of multiple clinico-pathological manifestations among a broad range of diseases; however, they provide relevant quantitative insights needed for standardization and diagnosis.
\end{itemize}

These challenges limit the translation from research to clinical diagnostics. We intend to give the readers some insights into the core problems behind the issues listed by briefly introducing WSI preparation and image acquisition protocols. Besides, we describe the state of the art of the proposed methods.
\subsection{Why AI for brain disorders?}

An important role of CP in brain disorders is related to the study and assessment of brain tumors as they cause significant morbidity and mortality worldwide, and pathology data is often available. In 2022 \cite{CancerNet_Editorial_Board2012-rt}, over 25k adults (14170 men and 10880 women) in the United States will have been diagnosed with primary cancerous tumors of the brain and spinal cord. 85\% to 90\% of all primary central nervous system (CNS) tumors (benign and cancerous) are located in the brain. Worldwide, over 300k people were diagnosed with a primary brain or spinal cord tumor in 2020. This disorder does not distinguish age, as nearly 4.2k children under the age of 15 will have also been diagnosed with brain or CNS tumors in 2022, in the United States. 

It is estimated that around one billion people have a mental or substance use disorder \cite{Ritchie2019-ju}. 
Some other key figures related to mental disorders worldwide are given by \cite{GBD_2017_DALYs_and_HALE_Collaborators2018-pm}. Globally, an estimated 264 million people are affected by depression. Bipolar disorder affects about 45 million people worldwide. Schizophrenia affects 20 million people worldwide, and approximately 50 million have dementia. In Europe, an estimated 10.5 million people have dementia, and this number is expected to increase to 18.7 million in 2050 \cite{European_Brain_Council2019-db}.


In the neurodegenerative disease group, 50 million people worldwide are living with Alzheimer's and other types of dementia \cite{Alzheimers_association2022-sy},  Alzheimer's disease being the underlying cause in 70\% of people with dementia \cite{European_Brain_Council2019-db}. Parkinson's disease affects approximately 6.2 million people worldwide \cite{GBD_2015_Neurological_Disorders_Collaborator_Group2017-bh} and represents the second most common neurodegenerative disorder. As the incidence of Alzheimer's and Parkinson's diseases rises significantly with age and people's life expectancy has increased, the prevalence of such disorders is set to rise dramatically in the future. For instance, there may be nearly 13 million people with Parkinson's by 2040 \cite{GBD_2015_Neurological_Disorders_Collaborator_Group2017-bh}.

Brain injuries are also the subject of a considerable number of incidents. Every year, around 17 million people suffer a stroke worldwide, with an estimate of 1 in 4 persons having a stroke during their lifetime \cite{Stroke_Alliance_for_Europe2016-rs}. Besides, stroke is the second cause of death worldwide and the first cause of acquired disability \cite{European_Brain_Council2019-db}.

These disorders also impact American regions, with over 500k deaths reported in 2019, due to neurological conditions. Among the conditions analyzed, the most common ones were Alzheimer's disease, Parkinson's, epilepsy, and multiple sclerosis \cite{Pan_American_Health_Organization2021-by}. 

In the case of brain tumors, treatment and prognosis require accurate and expedient histological diagnosis of the patient's tissue samples. Trained pathologists visually inspect histology slides, following a time-consuming and labor-intensive procedure. Therefore, the emergence of CP has triggered great hope to ease this tedious task and make it more robust. Clinical workflows in oncology rely on predictive and prognostic molecular biomarkers. However, the growing number of these complex biomarkers increases the cost and the time for decision-making in routine daily practice. Available tumor tissue contains an abundance of clinically relevant information that is currently not fully exploited, often requiring additional diagnostic material. Histopathological images contain rich phenotypic information that can be used to monitor underlying mechanisms contributing to disease progression and patient survival outcomes.

In most other brain diseases, histological images are only acquired post-mortem, and this procedure is far from being systematic. Indeed, it depends on the previous agreement of the patient to donate their brain for research purposes. Moreover, as mentioned above, the inspection of such images is complex and tedious, which further explains why it is performed in a minority of cases. Nevertheless, histopathological information is of the utmost importance in understanding the pathophysiology of most neurological disorders, and research progress would be impossible without such images. Finally, there are a few examples, beyond brain tumors, in which a surgical operation leads to an inspection of resected when the patient is alive (this is, for instance, the case of pharmacoresistant epilepsy).

Intraoperative decision-making also relies significantly on histological diagnosis, which is often established when a small specimen is sent for immediate interpretation by a neuropathologist. In poor resource settings, access to specialists may be limited, which has prompted several groups to develop Machine Learning (ML) algorithms for automated interpretation. Computerized analysis of digital pathology images offers the potential to improve clinical care (e.g., automated assistive diagnosis) and catalyze research (e.g., discovering disease subtypes or understanding the pathophysiology of a brain disorder).
\subsection{How do we present the information?}

In order to understand the potential and limitations of computational pathology algorithms, one needs to understand the basics behind the preparation of tissue samples and the image acquisition protocols followed by scanner manufacturers. Therefore, we have structured the chapter as follows.

Section \ref{sec:understading} presents an overview of tissue preservation techniques and how they may impact the final whole slide image. Section \ref{sec:histopath_img_analysis} introduces the notion of digital pathology and computational pathology, and its differences. It also develops the image acquisition protocol and describes the pyramidal structure of the WSI and its benefits. In addition, it discusses the possible impact of scanners on image processing algorithms. Section \ref{sec:methods} describes some of the state-of-the-art algorithms in artificial intelligence and its subcategories (machine learning and deep learning). This section is divided into methods for classifying and segmenting structures in WSI; and techniques that leverage deep learning algorithms to extract meaningful features from the WSI and apply them to a specific clinical application. Finally, Section \ref{sec:perspectives} explores new horizons in digital and computational pathology regarding explainability and new microscopic imaging modalities to improve tissue visualization and information retrieval.
\section{Understanding histological images}
\label{sec:understading}

We dedicate this section to understanding the process of acquiring histological images. We begin by introducing the two main tissue preservation techniques used in neuroscience studies, i.e., the routine-FFPE preparation (Formalin-Fixed Paraffin-Embedded) and the frozen tissue. We describe the process involved in each method and the main limitations for obtaining an appropriate histopathological image for analysis. Finally, we present the main procedures used in anatomopathology, based on such tissue preparations.

\subsection{Formalin-Fixed Paraffin-Embedded Tissue}

FFPE is a technique used for preserving biopsy specimens for clinical examination, diagnostic, experimental research, and drug development. A correct histological analysis of tissue morphology and biomarker localization in tissue samples will hinge on the ability to achieve high-quality preparation of tissue samples, which usually requires three critical stages: fixation, processing (also known as pre-embedding), and embedding.

Fixation is the process that allows the preservation of the tissue architecture (i.e., its cellular components, extracellular material, and molecular elements). Histotechnologists perform this procedure right after removing the tissue, in case of surgical pathology, or soon after death, during autopsy. Time is essential in preventing the autolysis and necrosis of excised tissues and preserving their antigenicity. Five categories of fixatives are used in this stage: aldehydes, mercurials, alcohols, oxidizing agents, and picrates. The most common fixative used for imaging purposes is formaldehyde (also known as formalin), included in the aldehydes group. Fixation protocols are not standardized and vary according to the type of tissue and the histologic details needed to analyze it. The variability in this stage induces the possibility for several factors to affect this process, such as buffering (pH regulation), penetration (also depending on tissue thickness), volume (the usual ratio is 10:1), temperature, fixative concentration (10\% solution is typical), and fixation time. These factors impact the quality of the scanned image, since stains used to highlight specific tissue properties may not react as expected.

After fixation, the tissue undergoes a processing stage necessary to create a paraffin embedding, which allows histotechnologists to cut the tissue into microscopic slides for further examination. The processing involves removing all water from the tissue using a series of alcohols and then clearing the tissue, which consists of removing the dehydrator with a miscible substance with the paraffin. Nowadays, tissue processors can automate this stage, by reducing inter-expert variability. 

\begin{figure}[hbtp]
	\centering
	\includegraphics[width=0.7\textwidth]{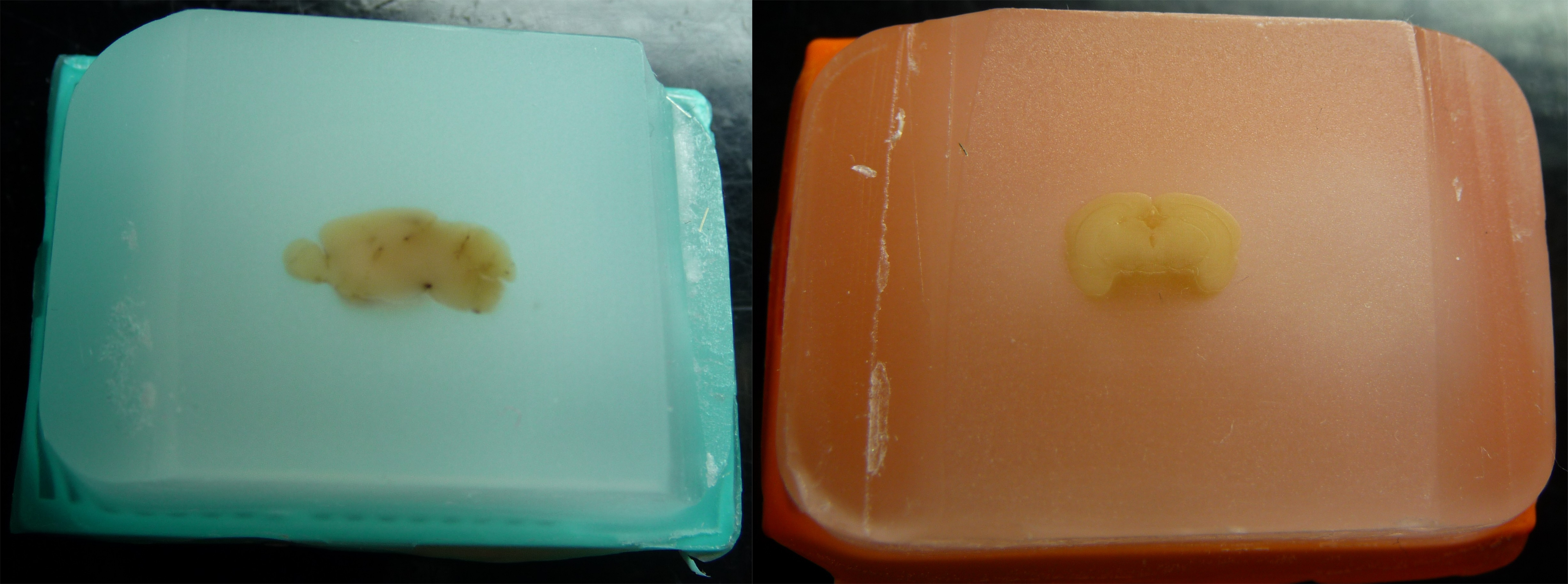}
	\caption{Paraffin casettes}
	\label{fig:parafin-cassette}
\end{figure}

Dehydration and clearing will leave the tissue ready for the technician to create the embedded paraffin blocks. Depending on the tissue, these embeddings must be correctly aligned and oriented, determining which tissue section or cut is studied. Also, the embedding parameters (e.g., embedding temperature or peculiar chemicals involved) may defer from the norm for unique studies, so the research entity and the laboratory making the acquisition need to define them beforehand. Figure \ref{fig:parafin-cassette} shows a paraffin embedding cassette where the FFPE tissue samples can be stored even at room temperature for long periods.

These embeddings undergo two more stages before being scanned: sectioning and staining. These procedures are discussed in the last section as they are no longer related to tissue preservation; instead, they are part of the tissue preparation stages before imaging. 

\subsection{Frozen Histological Tissue} 

Pathologists often use this tissue preservation method during surgical procedures where a rapid diagnosis of a pathological process is needed (extemporaneous preparation). In fact, frozen tissue produces the fastest stainable sections, although, compared to FPPE tissue, its morphological properties are not as good.

Frozen tissue (technically refer as cryosection) is created by submerging the fresh tissue sample into cold liquid (e.g., pre-cooled isopentane in liquid nitrogen) or by applying a technique called \textit{flash freezing}, which uses liquid nitrogen directly. As in FFPE, the tissue needs to be embedded in a medium to fix it to a chuck (i.e., specimen holder) in an optimal position for microscopic analysis. However, unlike FFPE tissue, no fixation or pre-embedding processes are needed for preservation.

For embedding, technicians use OCT (Optimal Cutting Temperature compound), a viscous aqueous solution of polyvinyl alcohol and polyethylene glycol designed to freeze, providing the ideal support for cutting the cryosections in the cryostat (microtome under cold temperature). Different embedding approaches exist depending on the tissue orientation, precision and speed of the process, tissue wastage, and the presence of freeze artifacts in the resulting image. Stephen R. Peters describes these procedures and other important considerations needed to prepare tissue samples using the frozen technique \cite{Peters2009-wr}. 

Frozen tissue preservation relies on storing the embeddings at low temperatures. Therefore, the tissue will degrade if the cold chain breaks due to tissue sample mishandling. However, as it better preserves the tissue's molecular genetic material, it is frequently used in sequencing analysis and immunohistochemistry (IHC). 

Other factors that affect the tissue quality and, therefore, the scanned images are the formation of ice crystals and the thickness of the sections. Ice crystals form when the tissue is not frozen rapidly enough, and it may negatively affect the tissue structure and, therefore, its morphological characteristics. On the other hand, frozen sections are often thicker than FFPE sections increasing the potential for lower resolution at higher magnifications and poorer images. 

\subsection{Tissue preparation}
We described the main pipeline to extract and preserve tissue samples for further analysis. Although the techniques described above can also be used for molecular and protein analysis (especially the frozen sections), we now focus only on the image pipeline by describing the slide preparation for scanning and the potential artifacts observed in the acquired images.

Once the tissue embeddings are obtained, either by FFPE or frozen technique, they are prepared for viewing under a microscope or scanner. The tissue blocks are cut, mounted on glass slides, and stained with pigments (e.g., Hematoxylin and Eosin [H\&E], Saffron, or molecular biomarkers) to enhance the contrast and highlight specific cellular structures under the microscope. 

Cutting the embeddings involves using a microtome to cut very thin tissue sections, later placed on the slide. The thickness of these sections is usually in the range of 4 to 20 microns. It will depend on the microscopy technique used for image acquisition and the experiment parameters. Special diamond knives are needed to get thinner sections, increasing the price of the microtome employed. If we use frozen embeddings, a cryostat keeps the environment's temperature low, avoiding tissue degradation. 

Once on the slide, the tissue is heated to adhere to the glass and avoid wrinkles. If warming the tissue damages some of its properties (especially for immunohistochemistry), glue-coated slides can be used instead. For cryosections, pathologists often prefer to add a fixation stage to resemble the readings of an FFPE tissue section. This immediate fixation is achieved using several chemicals, including ethanol, methanol, formalin, acetone, or a combination. S. Peters describes the differences in the image quality based on these fixatives, as well as the proposed protocol for cutting and staining frozen sections \cite{Peters2009-wr}. For FFPE sections, J. Zhang and H. Xiong \cite{Yuan2014-di} describe neural histology's cutting, mounting, and staining methods. Protocols suggested by the authors are valuable guidelines for histotechnologists as tissue usually folds or tears, and bubbles form when cutting the embeddings. Minimizing these issues is essential to have good-quality images and accurate quantification of histological results.

\begin{figure}[hbtp]
	\centering
		\includegraphics[width=0.9\textwidth]{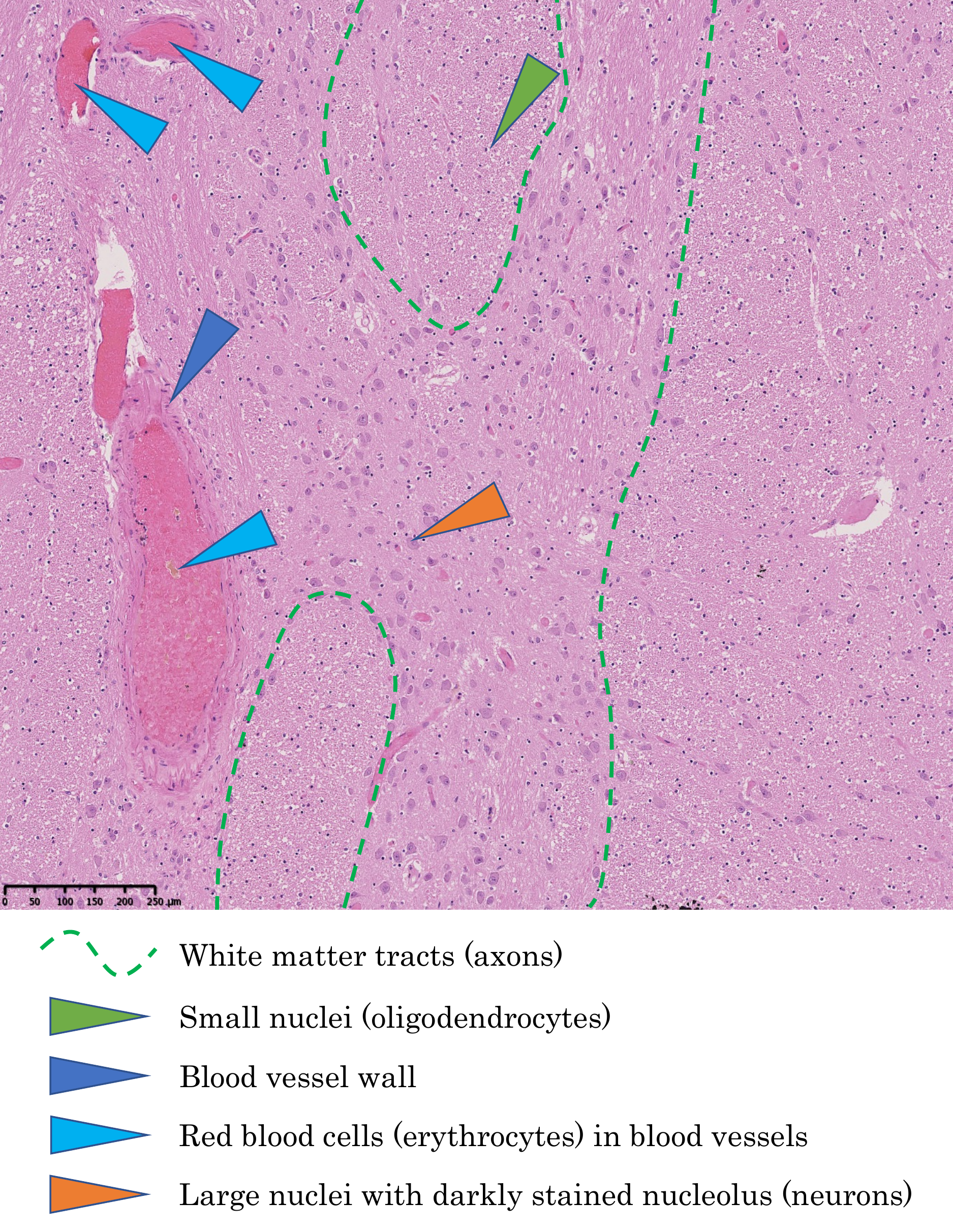}
	   \caption[H\&E-stained WSI]{H\&E-stained WSI from human brainstem tissue preserved using FFPE. Relevant structures were annotated by expert pathologist. {\sl Abbreviations. H\&E: Hematoxylin and Eosin. FFPE: Formalin-Fixed Paraffin-Embedded. WSI: Whole-Slide Image.}}
	\label{fig:HE-humanbrainstem}
\end{figure}

Staining is the last process applied to the tissue before being imaged. Staining agents do not react with the embedding chemicals used to preserve the tissue sample; therefore, the tissue section needs to be cleaned and dried beforehand (e.g., eliminating all remains of paraffin wax used in the embedding). In \cite{Titford2009-tq}, the authors present a review of the development of stains, techniques, and applications throughout time. One of the most common stains used in histopathology is Hematoxylin and Eosin (H\&E). This agent highlights cell nuclei with a purple-blue color and the extracellular matrix and cytoplasm with the characteristic pink. Other structures in the tissue will show different hues, shades, and combinations of these colors. Figure \ref{fig:HE-humanbrainstem} shows an H\&E stained human brainstem tissue and specific structures found on it. 

Other staining agents can be used depending on the structure we would like to study or the clinical procedure. For instance, the toluidine blue stain is frequently used for intraoperative consultation. Frozen sections are usually stained with this agent as it reacts almost instantly with the tissue. However, one disadvantage is that it only presents shades of blue and purple, so there is considerably less differential staining of the tissue structures \cite{Peters2009-wr}. 

\begin{figure}[hbtp]
	\centering
		\includegraphics[width=\textwidth]{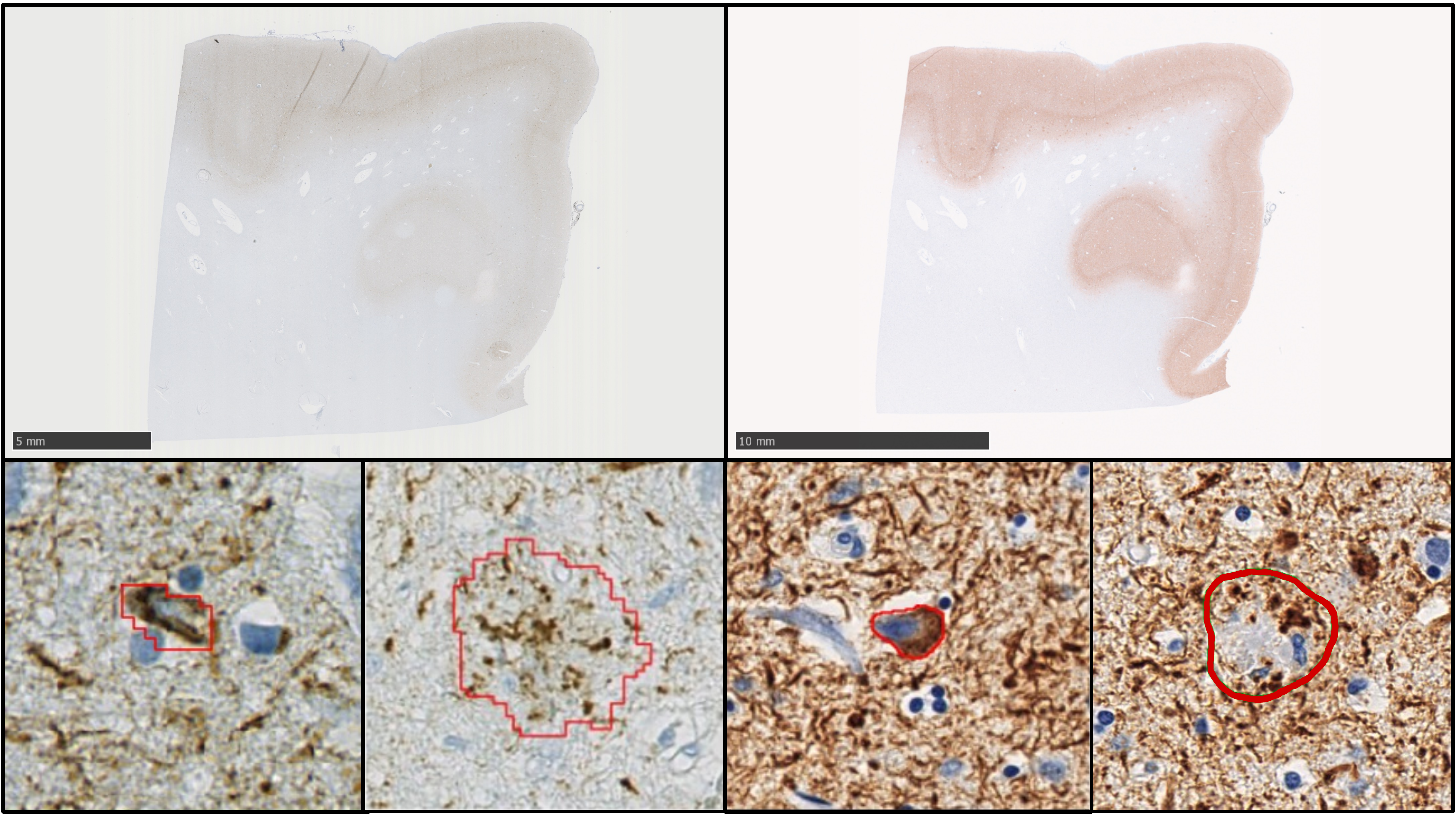}
	   \caption[Differences between ALZ50 and AT8 antibodies.]{[Top left] ALZ50 antibody used to discover compacted structures (tau pathologies). Below the WSI is an example of a neurofibrillary tangle (left) and a neuritic plaque (right) stained with ALZ50 antibody. [Top right] AT8 antibody, the most widely used in clinics, helps to discover all structures in a WSI. Below the WSI, there is an example of a neurofibrillary tangle (left) and a neuritic plaque stained with AT8 antibody (right). {\sl Abbreviation. WSI: Whole-Slide Image.}}
	\label{fig:ALZ50-AT8}
\end{figure}

For brain histopathology, other biomarkers are also available. For instance, the Cresyl Violet (or Nissl staining) is commonly used to identify the neuronal structure in the brain and spinal cord tissue \cite{Kapelsohn2015-pw}. Also, the Golgi method, which uses a silver staining technique, is used for observing neurons under the microscope \cite{Yuan2014-di}. Studies for Alzheimer's disease also frequently use ALZ50 and AT8 antibodies to reveal phosphorylated tau pathology using a standardized immunohistochemistry protocol \cite{Manouskova2022-hw, Jimenez2022-hs, Jimenez2022-hi}. Figure \ref{fig:ALZ50-AT8} shows the difference between ALZ50 and AT8 biomarkers and tau pathologies found in the tissue. 

Having the slide stained is the last stage to prepare for studying microscopic structures of diseased or abnormal tissues. Considering the number of people involved in these processes (pathologists, pathology assistants, histotechnologists, tissue technicians, and trained repository managing personnel) and the precision of each stage, standardizing certain practices to create valuable slides for further analysis is needed. 

Eiseman E. \textit{et al.} \cite{Eiseman2003-jk} reported a list of best practices for biospecimen collection, processing, annotation, storage, and distribution. The proposal aims to set guidelines for managing large biospecimen banks containing the tissue sample embeddings excised from different organs with different pathologies and demographic distributions.

More specific standardized procedures for tissue sampling and processing have also been reported. For instance, in 2012, the Society of Toxicologic Pathology charged a Nervous System Sampling Working Group with devising recommended practices to routinely screen the central nervous system (CNS) and peripheral nervous system (PNS) during nonclinical general toxicity studies. The authors proposed a series of approaches and recommendations for tissue fixation, collection, trimming, processing, histopathology examination, and reporting \cite{Bolon2013-tc}. Zhang J. \textit{et al.} also address the process of tissue preparation, sectioning, and staining but focus only on brain tissue \cite{Yuan2014-di}. Although these recommendations aim to standardize specific techniques among different laboratories, they are usually imprecise and approximate, leaving the specialists the final decision based on the tissue handled. 

\begin{figure}[hbtp]
	\centering
		\includegraphics[width=\textwidth]{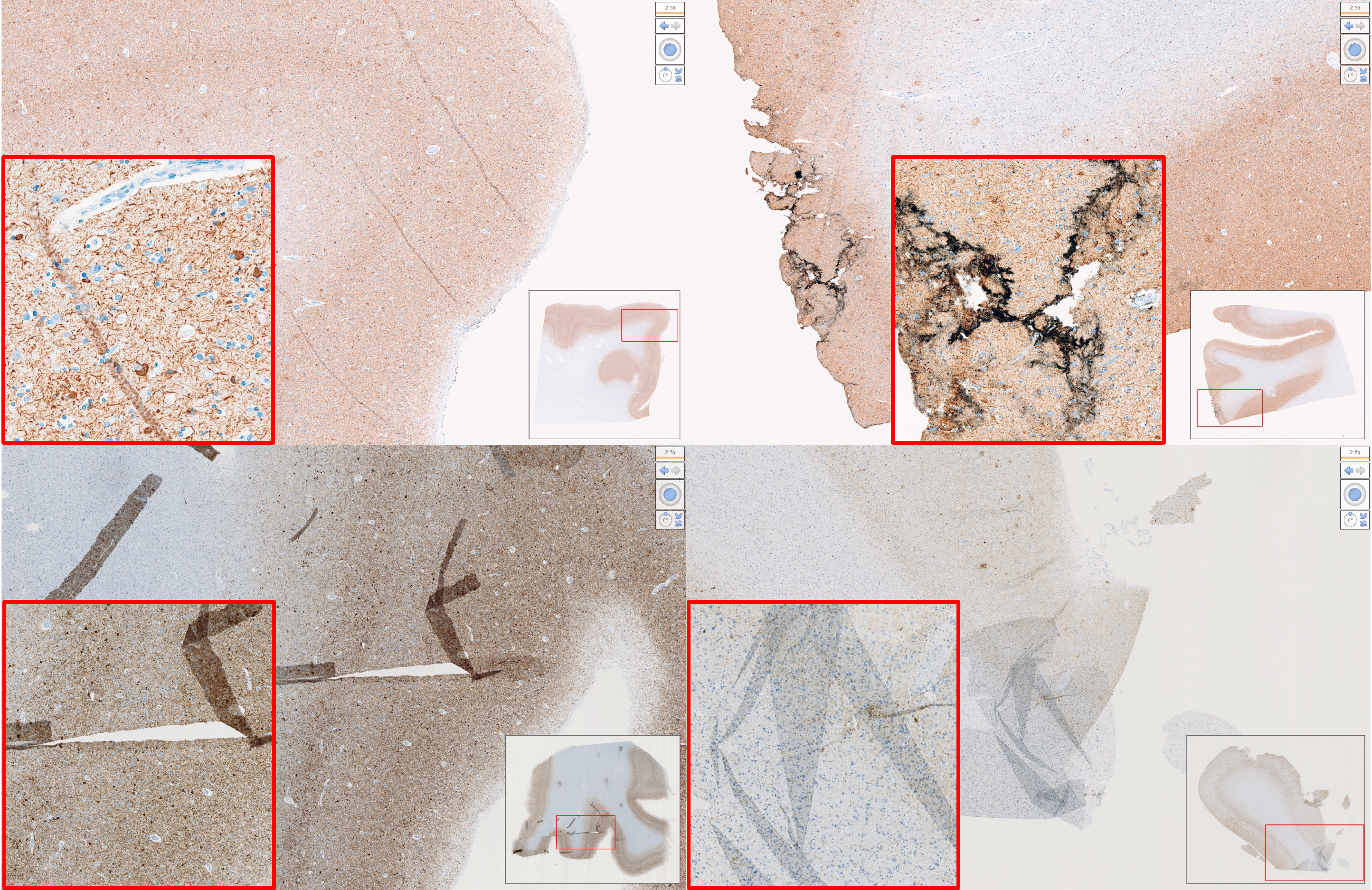}
	   \caption[Artifacts in WSI]{[Top left] Folding artifact (floatation and mounting related artifact), [Top right] Marking fixation process (fixation artifact), [Bottom left] Breaking artifact (microtome-related artifact), [Bottom right] Overlaying tissue (mounting artifact).}
	\label{fig:artifacts-WSI}
\end{figure}

Due to this lack of automation during surgical removal, fixation, tissue processing, embedding, microtomy, staining, and mounting procedures, several artifacts can impact the quality of the image and the results of the analysis. A review of these artifacts is presented in \cite{Taqi2018-bd}. The authors review the causes of the most frequent artifacts, how to identify them, and propose some ideas to prevent them from interfering with the diagnosis of lesions. For better understanding and following the tissue preparation and image acquisition procedure, the authors proposed a classification of eight classes: prefixation artifacts, fixation artifacts, artifacts related to bone tissue, tissue‑processing artifacts, artifacts related to microtomy, artifacts related to floatation and mounting, staining artifacts, and mounting artifacts. Figure \ref{fig:artifacts-WSI} shows some of them.

\section{Histopathological Image Analysis}
\label{sec:histopath_img_analysis}

This section aims to better understand the role that digital pathology plays in the analysis of complex and large amounts of information obtained from tissue specimens. As an additional option to incorporate more images with higher throughput, whole slide image scanners are briefly discussed. Therefore, we must discuss the DICOM standard used in medicine to digitally represent the images and, in this case, the tissue samples. We then focus on computational pathology, which is the analysis of the reconstructed whole slide images using different pattern recognition techniques such as machine learning (including deep learning) algorithms. This section contains some extractions from Jiménez's thesis work \cite{Jimenez_Garay2019-za}.

\subsection{Digital Pathology}
\label{ssec:dp}

Digital systems were introduced to the histopathological examination in order to deal with complex and vast amounts of information obtained from tissue specimens. Digital images were originally generated by mounting a camera on the microscope. The static pictures captured only reflected a small region of the glass slide, and the reconstruction of the whole glass slide was not frequently attempted due to its complexity and the fact that it is time-consuming. However, precision in the development of mechanical systems has made possible the construction of whole-slide digital scanners. Garcia \textit{et al.} \cite{Garcia_Rojo2006-oj} reviewed a series of mechanical and software systems used in the construction of such devices. The stored high-resolution images allow pathologists to view, manage, and analyze the digitized tissue on a computer monitor, similar to under an optical microscope but with additional digital tools to improve the diagnosis process.  

WSI technology, also referred to as \textit{virtual microscopy}, has proven to be helpful in a wide variety of applications in pathology (e.g., image archiving, telepathology, image analysis). In essence, a WSI scanner operation principle consists of moving the glass slide a small distance every time a picture is taken to capture the entire tissue sample. Every WSI scanner has six components: (a) a microscope with lens objectives, (b) a light source (bright field and/or fluorescent), (c) robotics to load and move glass slides around, (d) one or more digital cameras for capture, (e) a computer, and (f) software to manipulate, manage and view digital slides \cite{Farahani2015-ph}. The hardware and software used for these six components will determine the key features to analyze when choosing a scanner. Some research articles have compared the hardware and software capabilities of different scanners in the market. For instance, in \cite{Farahani2015-ph}, N. Farahani \textit{et al.} compared 11 WSI scanners from different manufacturers regarding imaging modality, slide capacity, scan speed, image magnification, image resolution, digital slide format, multilayer support, and special features their hardware and software may offer. This study showed that robotics and hardware used in a WSI scanner are currently state-of-the-art and almost standard in every device. Software, on the other hand, has some ground for further development. A similar study by Garcia \textit{et al.} \cite{Garcia_Rojo2006-oj} reviewed 31 digital slide systems comparing the same characteristics in Farahani's work. In addition, the authors classified the devices into digital microscopes (WSI) for virtual slide creation and diagnosis-aided systems for image analysis and telepathology. Automated microscopes were also included in the second group as they are the baseline for clinical applications. 

\subsection{Whole Slide Image structure}
\label{ssec:wsi_reconstr}

The Digital Imaging and Communications in Medicine (DICOM) standard was adopted to store WSI digital slides into commercially available PACS (Picture Archiving and Communication System) and facilitate the transition to digital pathology in clinics and laboratories. Due to the WSI dimension and size, a new pyramidal approach for data organization and access was proposed by the DICOM Standards Committee in \cite{DICOM_Standards_Committee2010-kk}.

\begin{figure}[ht!]
\centering
\includegraphics[width=\textwidth]{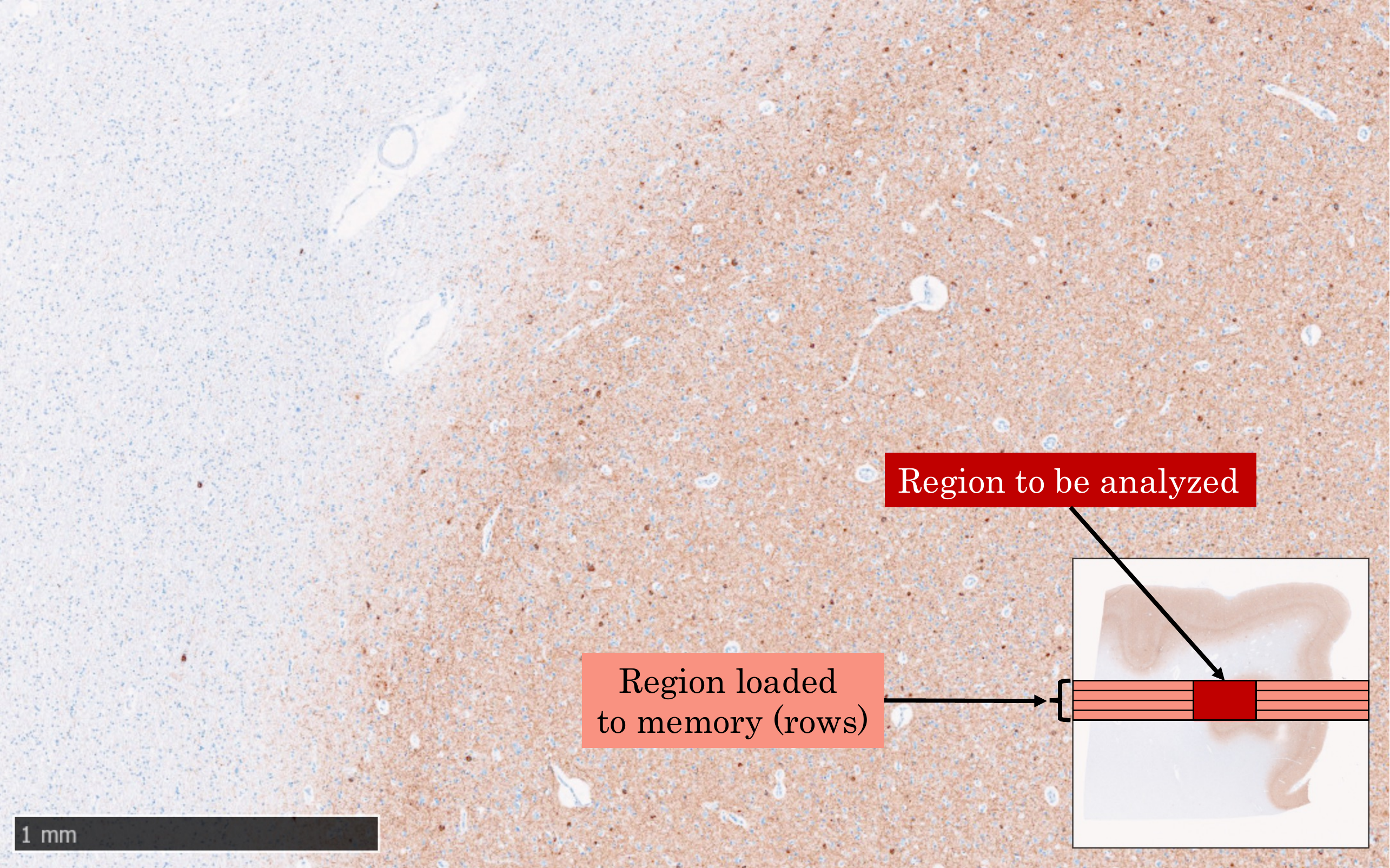}
\caption{\textit{Single frame} organization of Whole Slide Images.} 
\label{fig:DICOM1}
\end{figure}

A typical digitalization of a $20$mm $\times 15$mm sample using a resolution of $0.25\mu$m/pixel, also referred to as $40\times$ magnification, will generate an image of approximately $80000\times60000$ pixels. Considering a 24-bit color resolution, the digitized image size is about $15$GB. Data size might even go one order of magnitude higher if the scanner is configured to a higher resolution (e.g. $80\times$, $100\times$), Z planes are used, or additional spectral bands are also digitized. In any case, conventional storage and access to these images will demand excessive computational resources to be implemented into commercial systems. Figure \ref{fig:DICOM1} describes the traditional approach (i.e. \textit{single frame} organization), which stores the data in rows that extend across the entire image. This row-major approach has the disadvantage of loading unnecessary pixels into memory, especially if we want to visualize a small region of interest. 

\begin{figure}[ht!]
\centering
\includegraphics[width=\textwidth]{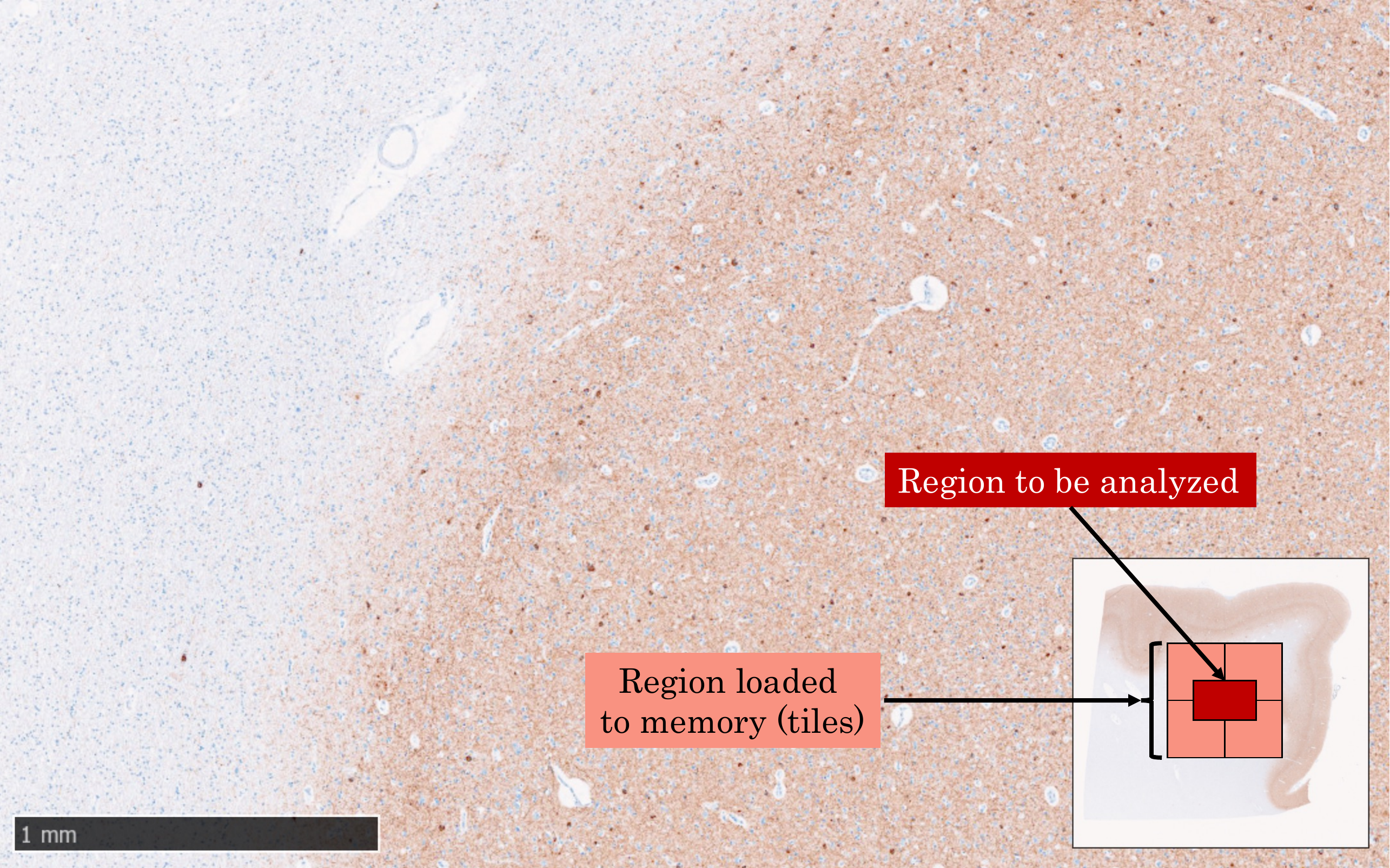}
\caption{Tiled image organization of Whole Slide Images. Tiles' size can range from 240 $\times$ 240 pixels up to 4096 $\times$ 4096 pixels.} 
\label{fig:DICOM2}
\end{figure}

\begin{figure}[ht!]
\centering
\includegraphics[width=\textwidth]{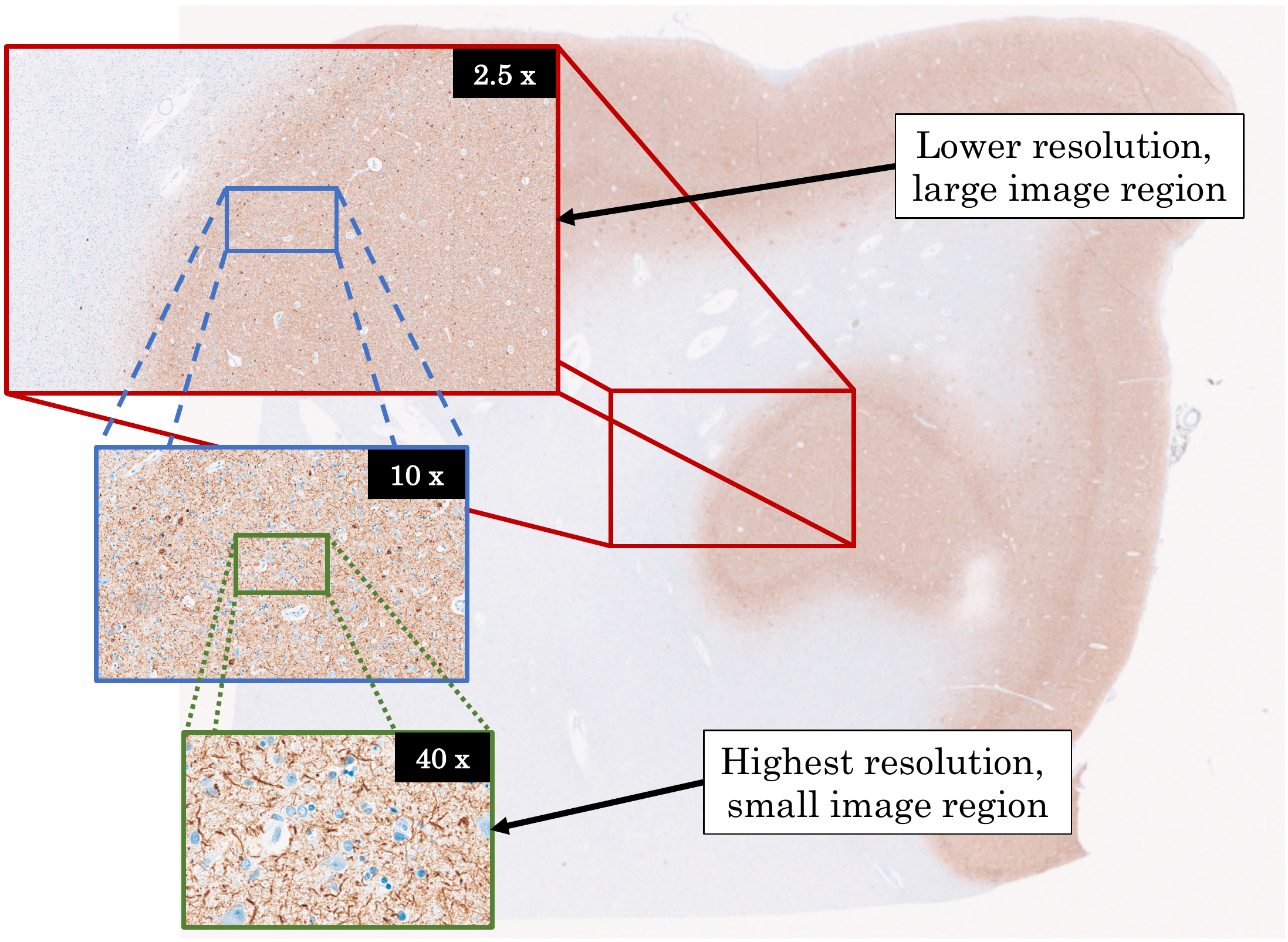}
\caption{Rapid zooming issue when accessing lower resolution images: large amount of data need to be loaded into memory. In this example, the image size at the highest resolution (221 nm/pixels) is 82432 $\times$ 80640 pixels.}
\label{fig:DICOM3}
\end{figure}

\begin{figure}[ht!]
\centering
\includegraphics[width=\textwidth]{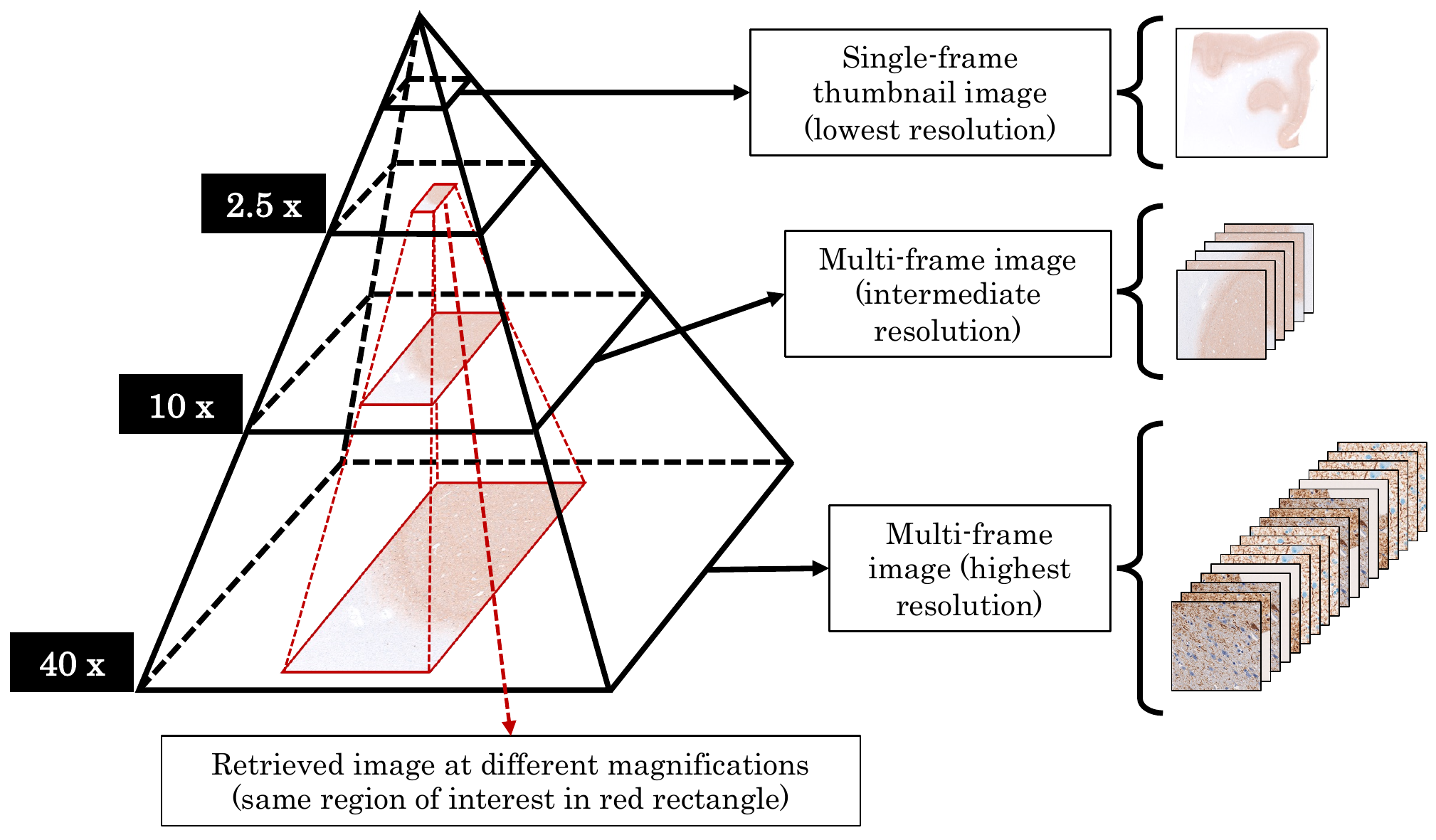}
\caption{Pyramidal organization of Whole Slide Images. In this example, the image size at the highest resolution (221 nm/pixels) is 82432 $\times$ 80640 pixels. The compressed (JPEG) file size is 2.22 GB, whereas the uncompressed version is 18.57 GB.}
\label{fig:DICOM4}
\end{figure}

Other types of organizations have also been studied. Figure \ref{fig:DICOM2} describes the storage of pixels in \textit{tiles}, which decreases the computational time for visualization and manipulation of WSI by loading only the subset of pixels needed into memory. Although this approach allows faster access and rapid visualization of the WSI, it fails when dealing with different magnifications of the images, as is the case in WSI scanners. Figure \ref{fig:DICOM3} depicts the issues with rapid zooming of WSI. Besides loading a larger subset of pixels into memory, algorithms to perform the down-sampling of the image are time-consuming. At the limit, to render a low-resolution thumbnail of the entire image, all the data scanned must be accessed and processed \cite{DICOM_Standards_Committee2010-kk}. Stacking precomputed low-resolution versions of the original image was proposed in order to overcome the zooming problem. Figure \ref{fig:DICOM4} describes the \textit{pyramidal} structure used to store different down-sampled versions of the original image. The bottom of the pyramid corresponds to the highest resolution and goes up to the thumbnail (lowest resolution) image. For further efficiency, tiling and pyramidal methods are combined to facilitate rapid retrieval of arbitrary subregions of the image as well as access to different resolutions. As depicted in Figure \ref{fig:DICOM4}, each image in the pyramid is stored as a series of tiles. In addition, the baseline image tiles can contain different color or z-planes if multispectral images are acquired or if tracking variations in the specimen thickness is needed. This combined approach can be easily integrated into a web architecture such as the one presented by Lajara \textit{et al.} \cite{Lajara2019-vm} as tiles of the current user's viewport can be cached without high memory impact.

As mentioned in previous paragraphs, WSI can occupy several terabytes of memory due to the data structure. Depending on the application, lossless or lossy compression algorithms can be applied. Lossless compression typically yields a 3X$-$5X reduction in size; meanwhile, lossy compression techniques such as JPEG and JPEG2000 can achieve from 15X$-$20X up to 30X$-$50X reduction respectively \cite{DICOM_Standards_Committee2010-kk}. Due to no standardization of WSI file formats, scan manufacturers may also develop their proprietary compression algorithms based on JPEG and JPEG2000 standards. Commercial WSI formats have a mean default compression value ranging from 13X to 27X. Although the size of WSI files is considerably reduced, efficient data storage was not the main issue when designing WSI formats for more than 10 years. In \cite{Helin2018-yz}, Helin \textit{et al.} addressed this issue and proposed an optimization to the JPEG2000 format, which yields up to 176X compression. Although no computational time has been reported in the aforementioned study, this breakthrough allows for efficient transmission of data through systems relying on internet communication protocols.

\subsection{Computational Pathology}
\label{ssec:compath}

Computational pathology is a term that refers to the integration of WSI technology and image analysis tools in order to perform tasks that were too cumbersome or even impossible to undertake manually. Image processing algorithms have evolved, yielding enough precision to be considered in clinical applications, such is the case for surgical pathology using frozen samples reported by W. Bauer \textit{et al.} in \cite{Bauer2015-zz}. Other examples mentioned in \cite{Farahani2015-ph} include morphological analysis to quantitatively measure histological structures \cite{Kong2013-sz}; automated selection of regions of interest such as areas of most active proliferative rate \cite{Lu2014-hl}; and automated grading of tumors \cite{Yeh2014-gg}. Moreover, educational activities have also benefited from the development of computational pathology. Virtual tutoring, online medical examinations, performance improvement programs, and even interactive \textit{illustrations} in articles and books are being implemented thanks to this technology \cite{Farahani2015-ph}. 

In order to validate a WSI scanner for clinical use (diagnosis purposes), several tests are conducted following the guidelines developed by the College of American Pathologists (CAP) \cite{Pantanowitz2013-tq}. On average, reported discrepancies between digital slides and glass slides are in the range of 1\% to 5\%. However, even glass-to-glass slide comparative studies can yield discrepancies due to observer variability, and increasing case difficulty.

Although several studies in the medical community have reported using WSI scanners to perform the analysis of tissue samples, pathologists remain reluctant to adopt this technology in their daily practice. Lack of training, limiting technology, shortcomings in scanning all slides, cost of equipment, and regulatory barriers have been reported as the principal issues \cite{Farahani2015-ph}. In fact, it was until early in 2017 that the first WSI scanner was approved by the FDA and released to the market \cite{Of_the_Commissioner2017-fi}. Nevertheless, WSI technology represents a milestone in modern pathology, having the potential to enhance the practice of pathology by introducing new tools which help pathologists provide a more accurate diagnosis based on quantitative information. Besides, this technology is also a bridge for bringing omics closer to routine histopathology towards future breakthroughs as spatial transcriptomics.
\section{Methods in Brain Computational Pathology} 
\label{sec:methods}

This section is dedicated to different machine learning and deep learning methodologies to analyze brain tissue samples. We describe the technology by focusing on how this is applied (i.e., at the WSI or the patch level), the medical task associated with it, the dataset used, the core structure/architecture of the algorithms, and the significant results. 

We begin by describing the general challenges in WSI analysis. Then we move on to deep learning methods concerning only WSI analysis, and we finalize with machine learning and deep learning applications for brain disorders focusing on the disease rather than the processing of the WSI. In addition, as in the primary biomedical areas, data annotation is a vital issue in computational pathology, generating accurate and robust results. Therefore, some new techniques used to create reliable annotations - based on a seed-annotated dataset will be presented and discussed.



\subsection{Challenges in WSI Analysis using ML}
Successful application of machine learning algorithms to WSIs can improve -- or even surpass -- the accuracy, reproducibility, and objectivity of current clinical approaches and propel the creation of new clinical tools providing new insights on various pathologies  \cite{Dimitriou2019-ij}. Due to the characteristics of a whole slide image and the acquisition process described in the sections above, researchers usually face two non-trivial challenges related to the visual understanding of the WSIs and the inability of hardware and software to facilitate learning from such high-dimensional images.

Regarding the first challenge, the issue relies on the lack of generalization of ML techniques due to image artifacts and color variability in staining. Imaging artifacts directly result from the tissue section processing errors and the hardware (scanners) used to digitize the slide. The uneven illumination, focusing, and image tiling are a few imaging artifacts present in the WSI, being the first the most relevant and studied as it is challenging for an algorithm to extract useful features from some regions of the scanned tissue. It gets even worse when staining artifacts such as stain variability are also present.

To address this problem, we find several algorithms for color normalization in the literature. Macenko \cite{Macenko2009-bn}, Vahadane \cite{Vahadane2015-ul}, Reinhard \cite{Reinhard2001-yu} are classical algorithms for color normalization implementing image processing techniques such as histogram normalization, color space transformations, color deconvolution (color unmixing), reference color density maps, or histogram matching. Extensions from these methods are also reported. For instance, Magee D. \textit{et al.} \cite{Magee2009-ex} proposed two approaches to extend the Reinhard method: a multimodal linear normalization in the \textit{Lab} colorspace and normalization in a representation space using stain-specific color deconvolution.

The use of machine learning techniques, specifically deep convolutional neural networks, has also been studied for color normalization. In \cite{Kang2021-ix}, the authors proposed the StainNet for stain normalization. The framework consists of a GAN\footnote{GAN: Generative Adversarial Networks} (\textit{teacher} network) trained to learn the mapping relationship between a source and target image; and an FCNN\footnote{FCNN: fully convolutional neural network} (\textit{student} network) able to transfer the mapping relationship of the GAN based on image content into a mapping relationship based on pixel values. A similar approach using cycle-consistent GANs was also proposed for the normalization of H\&E-stained WSIs \cite{Runz2021-iu}. In the last case, synthetically generated images capture the representative variability in the color space of the WSI, enabling the architecture to transfer any color information from a new source image into a target color space.

On the other hand, the second challenge related to the high dimensionality of WSIs is addressed in two ways: processing using patch-level or slide-level annotations. Dimitriou N. \textit{et al.} reported an overview of the literature for both approaches in \cite{Dimitriou2019-ij}. For patch-based annotations, the authors reported patch sizes ranging from 32 $\times$ 32 pixels up to 10000 $\times$ 10000 pixels and a frequent value of 256 $\times$ 256 pixels. Patches are generated and processed by sequentially dividing the WSI into tiles, which demand higher computational resources, by random sampling, leading to class imbalance issues, or by following a guided sampling based on pixel annotations. Patch-level annotations usually contain pixel-level labels. Frequently approaches using these annotations focus on the segmentation of morphological structures in patches rather than the classification of the entire WSI. In \cite{Jimenez2019-sn}, the authors studied the potential of semantic architectures such as the U-Net and compared it to classical CNN approaches for pixel-wise classification. Another approach known as HistoSegNet (\cite{Chan2019-zg}) implements a combination of visual attention maps (or activation maps) using the Grad-CAM algorithm and CNN for semantic segmentation of WSI. In addition, several methods are summarized in \cite{Ahmedt-Aristizabal2022-qy, Anklin2021-ww} using graph deep neural networks to detect and segment morphological structures in WSIs. 

Pixel labeling at high resolution is a time-demanding task and is prone to inter and intra-expert variabilities impacting the learning process of machine learning algorithms. Therefore, despite the lower granularity of labeling, several studies have shown promising results when working with slide-based annotations. 

With no available information about the pixel label, most algorithms usually aim to identify patches (or regions of interest in the WSI) that can collectively or independently predict the classification of the WSI. These techniques often rely on multiple instance learning, unsupervised learning, reinforcement learning, transfer learning, or a combination thereof \cite{Dimitriou2019-ij}. Tellez D. \textit{et al.} (\cite{Tellez2021-vq}) proposed a two-step method for gigapixel histopathology analysis based on an unsupervised neural network compression algorithm to extract latent representations of patches and a CNN to predict image-level labels from those compressed images. In \cite{Zhu2017-zi}, the authors proposed a 4-stage methodology for survival prediction based on randomly sampled patches from different patients' slides. They used PCA to reduce the features' space dimension prior to the K-means clustering process to group patches according to their phenotype. Then, a deep convolutional network (DeepConvSurv) is used to determine which patches are relevant for the aggregation and final survival score. Qaiser T. \textit{et al.} \cite{Qaiser2019-qd} proposed a model mimicking the histopathologist practice using recurrent neural networks (RNN) and CNN. In their proposal, they treat images as the \textit{environment} and the RNN+CNN as the \textit{agent} acting as a decision-maker (same as the histopathologists). The agent then looks at high-level tissue components (low magnification) and evaluates different regions of interest at low-level magnification, storing relevant morphological features into \textit{memory}. Similarly, Momeni A. \textit{et al.} \cite{Momeni2018-uy} suggested using deep recurrent attention models (DRAMs) and CNN to create an attention-based architecture to process large input patches and locate discriminatory regions more efficiently. This last approach needs, however, further validation as results are not conclusive and have not been accepted by the scientific community yet.

Relevant features for disease analysis, diagnosis, or patient stratification can be extracted from individual patches by looking into cell characteristics or morphology; however, higher structural information, such as the shape or extent of a tumor, can only be captured in more extensive regions. Some approaches to processing multiple magnification levels of a WSI are reported in \cite{Hou2016-cm,Campanella2018-su,Liu2017-sf,Van_Rijthoven2021-ij,Schmitz2021-mq}. They involve leveraging the pyramidal structure of WSI to access features from different resolutions and model spatial correlations between patches. 

All the studies cited so far have no specific domain of application. Most of them were trained and tested using synthetic or public datasets containing tissue pathologies from different body areas. Therefore, most of the approaches can extend to different pathologies and diseases. In the following subsections, however, we will focus only on specific brain disorders methodologies. 

\subsection{DL algorithms for Brain WSI analysis}

In recent times, deep-learning-based methods have shown promising results in digital pathology \cite{Janowczyk2016-xv}. Unfortunately, only a few public datasets contain WSI of brain tissue, and most of them only contain brain tumors. In addition, most of them are annotated at the slide level, making the semantic segmentation of structures more challenging. Independently of the task (i.e., detection/classification or segmentation) and the application in brain disorders, we will explore the main ideas behind the methodologies proposed in the literature.

For the analysis of benign or cancerous pathologies in brain tissue, tumor cell nuclei are of significant interest. The usual framework for analyzing such pathologies was reported in \cite{Kong2011-ny} and used the WSI of diffuse glioma. The method first segments the regions of interest by applying classical image processing techniques such as mathematical morphology and thresholding. Then, several handcrafted features such as nuclear morphometry, region texture, intensity, and gradient statistics were computed and inputted to a nuclei classifier. Although such an approach -- using Quadratic Discriminant Analysis and Maximum a Posteriori (MAP) as a classification mechanism -- reported an overall accuracy of 87.43\%, it falls short compared to CNN, which relies on automated feature extractions using convolutions rather than on handcrafted features. Xing F. \textit{et al.} \cite{Xing2016-hb} proposed an automatic learning-based framework for robust nucleus segmentation. The method begins by dividing the image into small regions using a sliding window technique. These patches are then fed to a CNN to output probability maps and generate initial contours for the nuclei using a region merging algorithm. The correct nucleus segmentation is obtained by alternating dictionary-based shape deformation and inference. This method outperformed classical image processing algorithms with promising results (mean Dice similarity coefficient of 0.85 and detection $F_1$ score of 0.77 computed using gold-standard regions within 15 pixels for every nucleus center) using CNN-based features over classical ones.

Following a similar approach, Xu Y. \textit{et al.} \cite{Xu2015-jh} reported the use of deep convolutional activation features for brain tumor classification and segmentation. The authors used a pre-trained AlexNet CNN \cite{Krizhevsky2012-zm} on the ImageNet dataset to extract patch features from the last hidden layer of the architecture. Features are then ranked based on the difference between the two classes of interest, and the top 100 are finally input to an SVM for classification. For the segmentation of necrotic tissue, an additional step involving probability mappings from SVM confidence scores and morphological smoothing is applied. Other approaches leveraging the use of CNN-based features for glioma are presented in \cite{Xu2017-kk,Hou2016-cm}. The experiments reported achieved a maximum accuracy of 97.5\% for classification and 84\% for segmentation. Although these results seemed promising, additional tests with different patch sizes in \cite{Hou2016-cm} suggested that the method's performance is data-dependent as numbers increase when larger patches, meaning more context information, are used. 

With the improvement of CNN architectures for natural images, more studies are also leveraging transfer learning to propose end-to-end methodologies for analyzing brain tumors. Ker \textit{et al.} (\cite{Ker2019-yx}) used a pre-trained Google Inception V3 network to classify brain histology specimens into normal, low-grade glioma (LGG) or high-grade glioma (HGG). Meanwhile, Truong \textit{et al.} (\cite{Truong2020-kz}) reported several optimization schemes for a pre-trained ResNet-18 for brain tumor grading. The authors also proposed an explainability tool base on tile-probability maps to aid pathologists in analyzing tumor heterogeneity. A summary of DL approaches used in brain WSI processing, alongside other brain imaging modalities such as MRI or CT, is reported by Zadeh \textit{et al.} in \cite{Zadeh_Shirazi2020-vf}.


Let us now focus on studies dealing with tau pathology, which is a hallmark of Alzheimer's disease. In \cite{Wurts2020-mh}, three different DL models were used to segment tau aggregates (tangles) and nuclei in post-mortem brain WSIs of patients with Alzheimer's disease. The 3 models included an FCNN, U-Net \cite{Ronneberger2015-ub}, and Segnet \cite{Badrinarayanan2017-zl}, with Segnet achieving the highest accuracy in terms of the intersection-over-union index. In \cite{Signaevsky2019-cz}, an FCNN was used on a dataset of 22 WSIs for semantic segmentation of tangle objects from post-mortem brain WSIs. Their model is able to segment tangles of varying morphologies with high accuracy under diverse staining intensities. An FCNN model is also used in \cite{Vega2021-pb} to classify morphologies of tau protein aggregates in the gray and white matter regions from 37 WSIs representing multiple degenerative diseases. In \cite{Manouskova2022-hw}, tau aggregate analysis is processed on a dataset of 6 post-mortem brain WSIs with a combined classification-segmentation framework which achieved an $F_1$ score of 81.3\% and 75.8\% on detection and segmentation tasks respectively. In \cite{Jimenez2022-hi}, neuritic plaques have been processed from 8 human brain WSI from the frontal lobe, stained with AT8 antibody (majorly used in clinics, helping to highlight most of the relevant structures). The impact of the staining (ALZ50 \cite{Manouskova2022-hw} vs. AT8 \cite{Jimenez2022-hi}), the normalization method, the slide scanner, the context, and the DL traceability/explainability have been studied, and a comparison with commercial software has been made. A baseline of 0.72 for the Dice score has been reported for plaque segmentation, reaching 0.75 using an attention U-Net.

Several domains in DL-based histopathological analysis of AD tauopathy remain unexplored. First, even if, as discussed, a first work concerning neuritic plaques has been recently published by our team in \cite{Jimenez2022-hi}, most of the existing works have used DL for segmentation of tangles rather than plaques, as the latter are harder to identify against the background gray matter due to their diffuse/sparse appearance. Secondly, annotations of whole slide images are frequently affected by errors by human annotators. In such cases, a  DL preliminary model may be trained using weakly annotated data and used to assist the expert in refining annotations. Thirdly, contemporary tau segmentation studies do not consider context information. This is important in segmenting plaques from brain WSIs as these occur as sparse objects against an extended background of gray matter. Finally, DL models with explainability features have not yet been applied in tau segmentation from WSIs. This is a critical requirement for DL models used in clinical applications \cite{Border2021-wh} \cite{Yamamoto2019-gk}. The DL models should not only be able to precisely identify regions of interest, but clinicians and general users need to know the discriminative image features the model identifies as influencing their decisions. 

\subsection{Applications of brain computational pathology}
Digital systems were introduced to the histopathological examination to deal with complex and vast amounts of information obtained from tissue specimens. Whole slide imaging technology has proven to be helpful in a wide variety of applications in pathology (e.g., image archiving, telepathology, image analysis), especially when combining this imaging technique with powerful machine learning algorithms (i.e., computational pathology). 

In this section, we will describe some applications of computational pathology for the analysis of brain tissue. Most methods focus on tumor analysis and cancer; however, we also find interesting results in clinical applications, drug trials (\cite{Lahiani2020-ld}), and neurodegenerative diseases. The authors cited in this section aim to understand brain disorders and use deep learning algorithms to extract relevant information from WSI. 

In brain tumor research, an early survival study for brain glioma is presented in \cite{Zhu2017-zi}. The approach has been previously described above. In brief, it is a 4-stage methodology based on randomly sampled patches from different patients' slides. They perform dimensionality reduction using PCA and then K-means clustering to group patches according to their phenotype. Then, the patches are sent to a deep convolutional network (DeepConvSurv) to determine which were relevant for the aggregation and final survival score. The deep survival model is trained on a small dataset leveraging the architecture the authors proposed. Also, the method is annotation free, and it can learn information about one patient, regardless of the number or size of the WSIs. However, it has a high computational memory footprint as it needs hundreds of patches from a single patient's WSI. In addition, the authors do not address the evaluation of the progression of the tumor, and a deeper analysis of the clusters could provide information about the phenotypes and their relation to brain glioma. 

Whole slide images have been used as a primary source of information for cancer diagnosis and prognosis, as they reveal the effects of cancer onset and its progression at the sub-cellular level. However, being an invasive image modality (i.e., tissue gathered during a biopsy), it is less frequently used in research and clinical settings. As an alternative, non-invasive and non-ionizing imaging modalities, such as MRI, are quite popular for oncology imaging studies, especially in brain tumors. 

Although radiology and pathology capture morphologic data at different biological scales, a combination of image modalities can improve image-based analysis. In \cite{Kurc2020-xx}, the authors presented three classification methods to categorize adult diffuse glioma cases into oligodendroglioma and astrocytoma classes using radiographic and histologic image data. Thirty-two cases were gathered from the TCGA project \footnote{https://www.cancer.gov/about-nci/organization/ccg/research/structural-genomics/tcga/using-tcga/typesan} containing a set of MRI data (T1, T1C, FLAIR, and T2 images) and its corresponding WSI, taken from the same patient at the same time point. The methods described were proposed in the context of the Computational Precision Medicine (CPM) satellite event at MICCAI 2018, one of the first combining radiology and histology imaging analyses. The first one develops two independent pipelines giving two probability scores for the prediction of each case. The MRI pipeline preprocesses all images to remove the skull, co-register, and re-sample the data to leverage a fully convolutional neural network (CNN) trained on another MRI dataset (i.e., BraTS-2018) to segment tumoral regions. Several radiomic features are computed from such regions, and after reducing its dimensionality with PCA, a logistic regression classifier outputs the first probability score. WSIs also need a preprocessing stage as tissue samples may contain large areas of glass background. After a color space transformation to HSV (hue saturation value), lower and upper thresholds are applied to get a binary mask with the region of interest, which is then refined using mathematical morphology. Color-normalized patches of $224 \times 224$ pixels are extracted from the region-of-interest (ROI) and filtered to exclude outliers. The remaining patches are used to refine a CNN (i.e., DenseNet-161) pre-trained on the ImageNet dataset. In the prediction phase, the probability score of the WSI is computed using a voting system of the classes predicted for individual patches. The scores from both pipelines are finally processed in a confidence-based voting system to determine the final class of each case. This proposal achieved an accuracy score of 0.9 for classification.

The second and third approaches also processed data in two different pipelines. There are slight variations in the WSI preprocessing step in the second method, including Otsu thresholding for glass background removal and histogram equalization for color normalization of patches of $448 \times 448$ pixels. Furthermore, the authors used a 3D CNN to generate the output predictions for the MRI data and a DenseNet pre-trained architecture for WSI patch classification. The last feature layer from each classification model is finally used as input to an SVM model for a unified prediction. In addition, regularization using dropout is performed in the test phase to avoid overfitting the models. The accuracy obtained with this methodology was 0.8.

The third approach uses larger patches from WSI and an active learning algorithm proposed in \cite{Qi2019-pr} to extract regions of interest instead of randomly sampling the tissue samples. Features from the WSI patches are extracted using a VGG16 CNN architecture. The probability score is combined with the output probability of a U-Net + 2D DenseNet architecture used to process the MRI data. The method achieved an accuracy of 0.75 for unified classification. Although results are promising and provide a valid approach to combining imaging modalities, data quality and quantity are still challenging. The use of pre-trained CNN architectures for transfer learning using a completely different type of imaging modality might impact the performance of the whole pipeline. As seen in previous sections, WSI presents specific characteristics depending on the preparation and acquisition procedures not represented in the ImageNet dataset. 

An extension to the previous study is presented in \cite{Wang2022-br}. The authors proposed a two-stage model to classify gliomas into three subtypes. WSIs were divided into tiles and filtered to exclude patches containing glass backgrounds. An ensemble learning framework based on three CNN architectures (EfficientNet-B2, EfficientNet-B3, and SEResNeXt101) is used to extract features which are then combined with meta-data (i.e., age of the patient) to predict the class of glioma. MRI data is preprocessed in the same way as described before and input to a 3D CNN network with a 3D ResNet architecture as a backbone. 

The release of new challenges and datasets, such as the Computational Precision Medicine: Radiology-Pathology Challenge on brain tumor classification (CPM-RadPath), has also allowed studies using weakly supervised deep learning methods for glioma subtype classification. For instance, in \cite{Hsu2022-dv}, the authors combine 2D and 3D CNN to process 388 WSI, and its corresponding multiparametric MRI collected from the same patients. Based on a confidence index, the authors were able to fuse WSI-and-MRI-based predictions improving the final classification of the glioma subtype.

Moving on from brain tumors, examining brain WSI also provides essential insights into spatial characteristics helpful in understanding brain disorders. 

In this area, analyzing small structures present in postmortem brain tissue is crucial to understanding the disease deeply. For instance, in Alzheimer's disease, tau proteins are essential markers presenting the best histopathological correlation with clinical symptoms \cite{Duyckaerts2009-hh}. Moreover, these proteins can aggregate in three different structures within the brain (i.e., neurites, tangles, and neuritic plaques) and constitute one significant biomarker to study the progression of the disease and stratify patients accordingly. 

In \cite{Manouskova2022-hw}, the authors addressed the detection task of the Alzheimer's patient stratification pipeline. The authors proposed a U-Net-based methodology for tauopathies segmentation and a CNN-based architecture for tau aggregates classification. In addition, the pipelines were completed with a non-linear color normalization preprocessing and a morphological analysis of segmented objects. These morphological features can aid in the clustering of patients having different disease manifestations. One limitation, however, is the accuracy obtained in the segmentation/detection process. 

Understanding the accumulation of abnormal tau protein in neurons and glia allows differentiating 
tauopathies such as Alzheimer's disease, progressive supranuclear palsy (PSP), cortico-basal degeneration (CBD), and Pick's disease (PiD). In \cite{Koga2022-nw}, the authors proposed a diagnostic tool consisting of two stages: (1) an object detection pipeline based on the CNN YOLOv3 and (2) a random forest classifier. The goal is to detect different tau lesion types and then analyze their characteristics to determine to which specific pathology they belong. With an accuracy of 0.97 over 2522 WSI, the study suggests that machine learning methods can be applied to help differentiate uncommon neurodegenerative tauopathies.

Tauopathies are analyzed using postmortem brain tissue samples. For \textit{in-vivo} studies, there exist tau PET tracers that, unfortunately, have not been validated and approved for clinical use as correlations with histological samples are needed. In \cite{Ushizima2022-pp}, the authors proposed an end-to-end solution for performing large-scale, voxel-to-voxel correlations between PET and high-resolution histological signals using open-source resources and MRI as the common registration space. A U-Net-based architecture segments tau proteins in WSI to generate 3D tau inclusion density maps later registered to MRI to validate the PET tracers. Although segmentation performance was around 0.91 accurate in 500 WSI, the most significant limitation is the tissue sample preparation, meaning extracting and cutting brain samples to reconstruct 3D histological volumes. Additional studies combining postmortem MRI and WSI for neurodegenerative diseases were reported by Laura Jonkman \textit{et al.,} in \cite{Jonkman2019-pl}.
\section{Perspectives}
\label{sec:perspectives}

This last section of the chapter deals with new techniques for the explainability of artificial intelligence algorithms. It also describes new ideas related to responsible artificial intelligence in the context of medical applications, computational histopathology, and brain disorders. Besides, it introduces new image acquisition technology mixing bright light and chemistry to improve intraoperative applications. Finally, we will highlight computational pathology's strategic role in spatial transcriptomics and refined personalized medicine.

\begin{figure}[!htb]
    \centering
    \includegraphics[width=\textwidth]{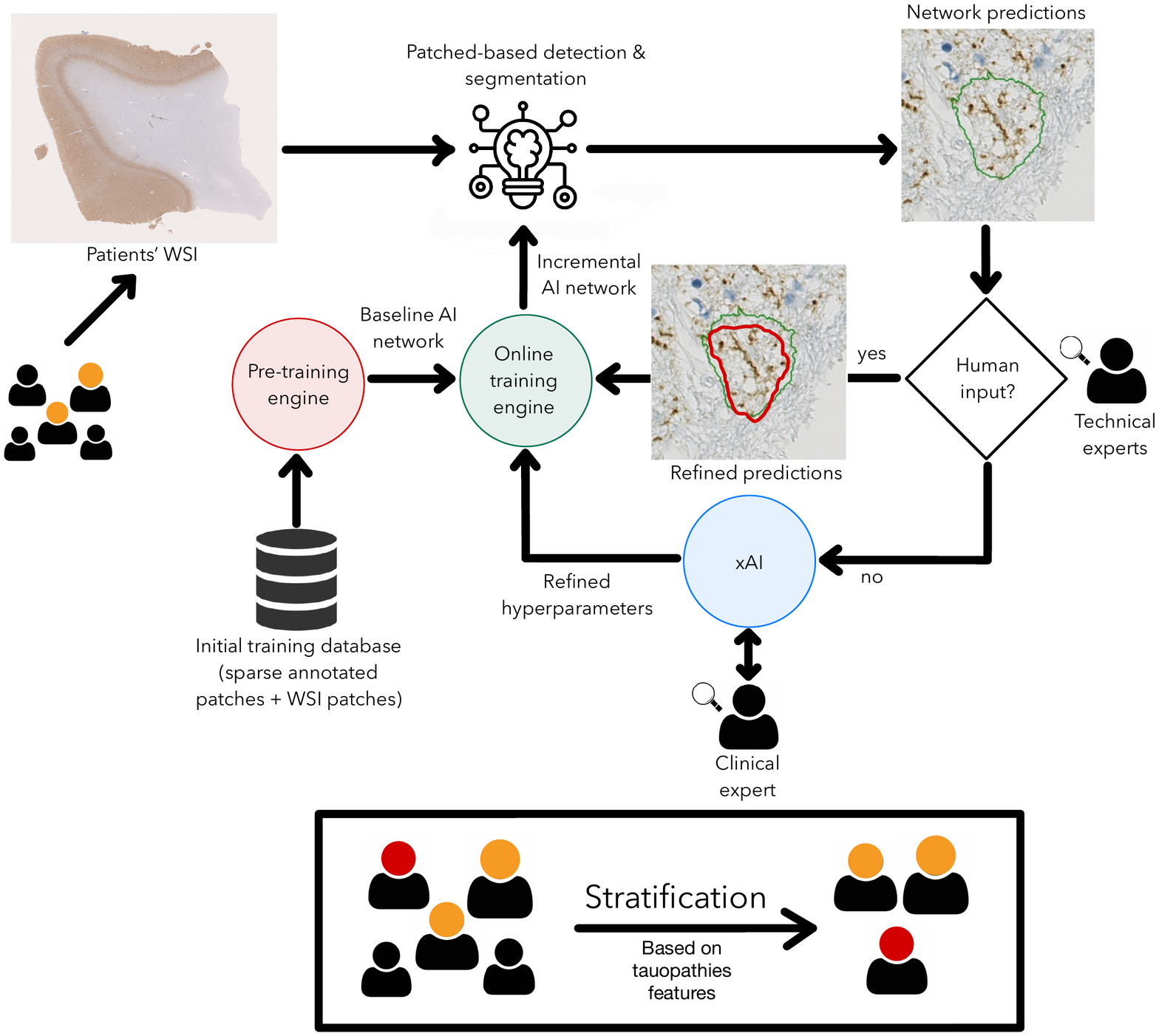}
    \caption{Expert-in-the-loop architecture proposal to improve tauopathies segmentation and to stratify AD patients.}
    \label{fig:human-in-the-loop}
\end{figure}

In \cite{Jimenez2022-hs, Jimenez2022-hi}, we address the issue of accurate segmentation by proposing a two-loop scheme as shown in Figure \ref{fig:human-in-the-loop}. In our method, a U-Net-based neural network is trained on several WSIs manually annotated by expert pathologists. The structures we focus on are neuritic plaques and tangles following the study in \cite{Manouskova2022-hw}. The network's predictions (in new WSIs) are then reviewed by an expert who can refine the predictions by modifying the segmentation outline or validating new structures found in the WSI. Additionally, an attention-based architecture is used to create a visual explanation and refine the hyperparameters of the initial architecture in charge of the prediction proposal. 

\begin{figure}[!htb]
    \centering
    \includegraphics[width=\textwidth]{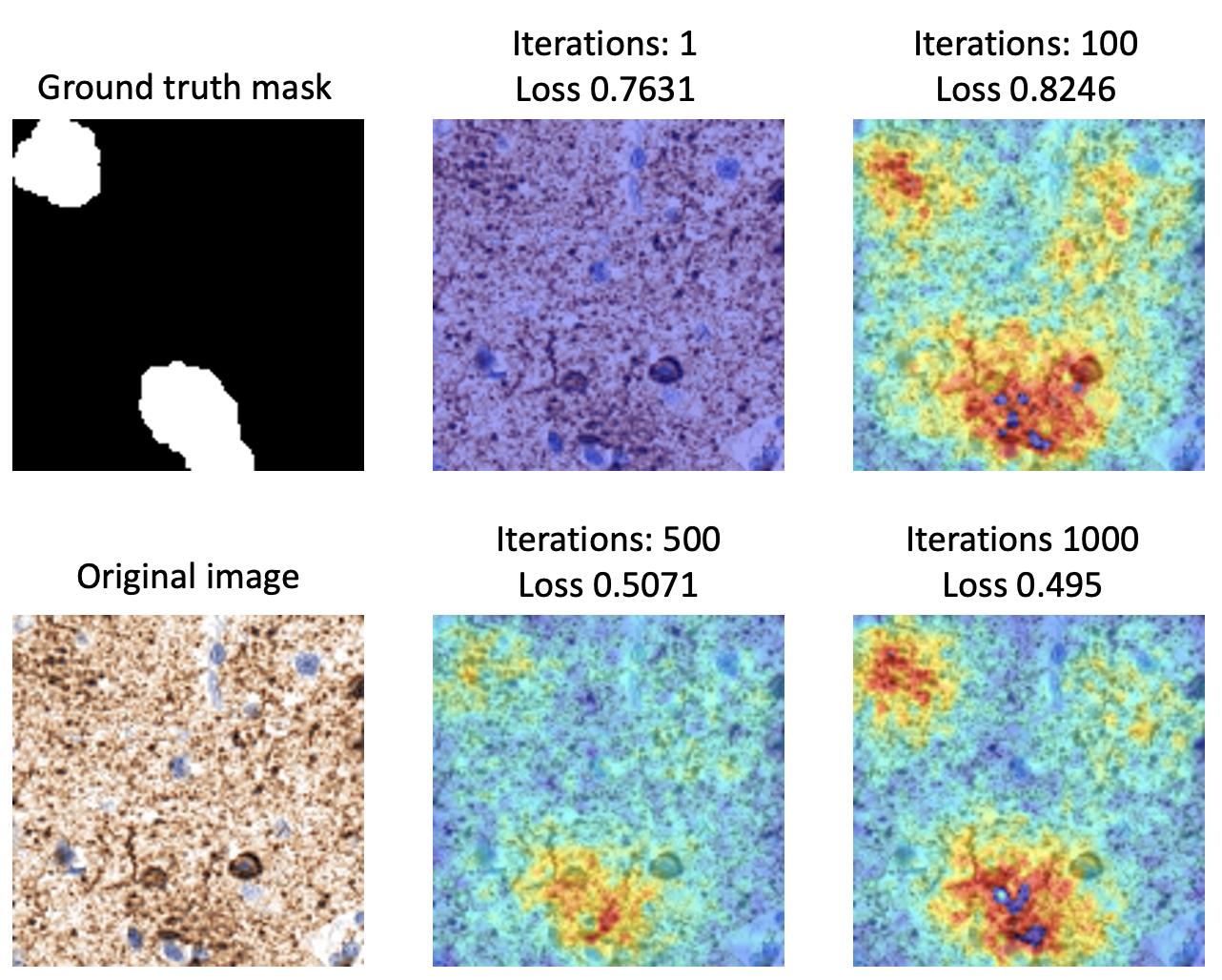}
    \caption{Attention U-Net results. The figure shows a patch of size $128 \times 128$ pixels, the ground-truth binary mask, and the focus progression using successive activation layers of the network.}
    \label{fig:attUnet}
\end{figure}

We tested the attention-based architecture with a dataset of 8 WSI divided into patches following an ROI-guided sampling. Results show qualitatively in Figure \ref{fig:attUnet} that through this visual explanation, the expert in the loop could define the border of the neuritic plaque (object of interest) more accurately so the network can update its weights accordingly. Additionally, quantitative results (dice score of approximately 0.7) show great promise for this attention U-Net architecture. 

Our next step is to use a single architecture for explainability and segmentation/classification. We believe our method will improve the accuracy of the neuritic plaques and tangles outline and create better morphological features for patient stratification and understanding of Alzheimer's Disease \cite{Jimenez2022-hs, Jimenez2022-hi}.

Despite their high computational efficiency, artificial intelligence -- in particular deep learning -- models face important usability and translational limitations in clinical use, as in biomedical research. The main reason for these limitations is generally low acceptability by biomedical experts, essentially due to the lack of feedback, traceability, and interpretability. Indeed, domain experts usually feel frustrated by a general lack of insights, while the implementation of the tool itself requires them to make a considerable effort to formalize, verify and provide a tremendous amount of domain expertise. Some authors speak of a ``black-box'' phenomenon, which is undesirable for a traceable, interpretable, explicable, and, ultimately, responsible use of these tools. 

In recent years, eXplainable AI (xAI) models have been developed to provide insights from and understand the AI decision-making processes by interpreting their second-opinion quantifications, diagnoses, and predictions. Indeed, while explaining simple AI models for regression and classification tasks is relatively straightforward, the explainability task becomes more difficult as the model's complexity increases. Therefore, a novel paradigm becomes necessary for better interaction between computer scientists, biologists, and clinicians, with the support of an essential new actor: xAI, thus opening the way towards Responsible AI: fairness, ethics, privacy, traceability, accountability, safety, and carbon footprint.

In digital histopathology, several studies report on the usage and the benefits of explainable AI models. In \cite{Tosun2020-aa}, the authors describe an xAI-based software named HistoMapr and its application to breast core biopsies. This software automatically identifies the regions of interest (ROI) and rapidly discovers key diagnostic areas from whole slide images of breast cancer biopsies. It generates a provisional diagnosis based on the automatic detection and classification of relevant ROIs and also provides a list of key findings to pathologists that led to the recommendation. An explainable segmentation pipeline for whole slide images is described in \cite{Chan2019-zg}, which does a patch-level classification of colon glands for different cancer grades using a CNN followed by inference of class activation maps for the classifier. The activation maps are used for final pixel-level segmentation. The method outperforms other weakly supervised methods applied to these types of images and generalizes to other datasets easily. A medical use-case of AI versus human interpretation of histopathology data using a liver biopsy dataset is described in \cite{Holzinger2019-jt}, which also stresses the need to develop methods for causability or measurement of the quality of AI explanations. In \cite{Yamamoto2019-gk},  AI models like deep auto-encoders were used to generate features from whole-mount prostate cancer pathology images that pathologists could understand. This work showed that a combination of human and AI-generated features produced higher accuracy in predicting prostate cancer recurrence. Finally, in \cite{Jimenez2022-hi}, the authors show that, besides providing valuable visual explanation insights, the use of attention U-Net is even helping to increase the results of neuritic plaques segmentation by pulling up the Dice score to 0.75 from 0.72 (with the original U-Net). 

Based on the fusion of MRI and histopathology imaging datasets, a deep learning 3D U-Net model with explanations is used in \cite{Gunashekar2022-ru} for prostate tumor segmentation. Grad‐CAM \cite{Selvaraju2020-yu} heat maps were estimated for the last convolutional layer of the U-Net for interpreting the recognition and localization capability of the U-Net. In \cite{Zeineldin2022-kg}, a framework named NeuroXAI is proposed to render explainability to existing deep learning models in brain imaging research without any architecture modification or reduction in performance. This framework implements seven state-of-the-art explanation methods - including Vanilla gradient \cite{Simonyan2014-zj}, Guided back-propagation, Integrated gradients \cite{Sundararajan2017-hy}, SmoothGrad \cite{Smilkov2017-rn} and Grad-CAM. These methods can be used to generate visual explainability maps for deep learning models like 2D and 3D CNN, VGG \cite{Simonyan2014-rz}, and Resnet-50 \cite{He2016-fh} (for classification) and 2D/3D U-Net (for segmentation). In \cite{Esmaeili2021-ou}, the high-level features of three deep convolutional neural networks (DenseNet-121, GoogLeNet, MobileNet) are analysed using the Grad-CAM explainability technique. The Grad-CAM outputs helped distinguish these three models' brain tumor lesion localization capabilities. An explainability framework using SHAP\cite{Lundberg2017-jr} and LIME\cite{Ribeiro2016-ok} to predict patient age using the morphological features from a brain MRI dataset is developed in \cite{Lombardi2021-yl}. The SHAP explainability model is robust for this imaging modality to explain morphological feature contributions in predicting age, which would ultimately help develop personalized age-related bio-markers from MRI. Attempts to explain the functional organization of deep segmentation models like DenseUnet, ResUnet, and SimUnet and understand how these networks achieve high accuracy brain tumor segmentation are presented in \cite{Natekar2020-tp}. While current xAI methods mainly focus on explaining models on single image modality, the authors of \cite{Jin2022-hj} address the explainability issue in multi-modal medical images, such as PET-CT or multi-stained pathological images. Combining modality-specific information to explain diagnosis is a complex clinical task, and the authors developed a new multi-modal explanation method with modality-specific feature importance.

Intraoperative tissue diagnostic methods have remained unchanged for over 100 years in surgical oncology. Standard light microscopy used in combination with H\&E and other staining biomarkers has improved over the last decades with the appearance of new scanner technology. However, the steps involved in the preparation and some artifacts introduced by scanners pose a potential barrier to efficient, reproducible, and accurate intraoperative cancer diagnosis and other brain disorder analyses. As an alternative, label-free optical imaging methods have been developed. 

Label-free imaging is a method for cell visualization which does not require labeling or altering the tissue in any way. Brightfield, phase contrast, and differential interference contrast microscopy can be used to visualize label-free cells. The two latter techniques are used to improve the image quality of standard brightfield microscopy. Among its benefits, the cells are analyzed in their unperturbed state, so findings are more reliable and biologically relevant. Also, it is a cheaper and quicker technique as tissue does not need any genetic modification or alteration. In addition, experiments can run longer, making them appropriate for studying cellular dynamics \cite{Bleloch2020-tu}. Raman microscopy, a label-free imaging technique, uses infrared incident light from lasers to capture vibrational signatures of chemical bonds in the tissue sample's molecules. The biomedical tissue is excited with a dual-wavelength fiber laser set up at the so-called \textit{pump} and \textit{Stokes} frequencies to enhance the weak vibrational effect \cite{Marx2019-vf}. This technique is known as coherent anti-Stokes Raman scattering (CARS) or stimulated Raman scattering histology (SRH). 

Sarri et al. \cite{Sarri2019-up} proposed the first one-to-one comparison between SRH and H\&E as the latter technique remains the standard in histopathology analyses. The evaluation was conducted using the same cryogenic tissue sample. SRH data was first collected as it did not need staining. SRH and SHG (second harmonic generation, another label-free non-linear optical technique) were combined to generate a virtual H\&E slide for comparison. The results evidenced the almost perfect similarity between SRH and standard H\&E slides. Both virtual and real slides show the relevant structures needed to identify cancerous and healthy tissue. In addition, SRH proved to be a fast histologic imaging method suitable for intraoperative procedures. 

Similar to standard histopathology, computational methods are also applicable to SRH technology. For instance, Todd C. Hollon et al. \cite{Hollon2020-iq} proposed a CNN methodology to interpret histologic features from SRH brain tumor images and accurately segment cancerous regions. Results show a slightly better performance (94.6\%) than the one obtained by the pathologist (93.9\%) in the control group. This study was extended and validated for intraoperative diagnosis in \cite{Hollon2020-um}. The study used 2.5 million SRH images and predicted brain tumor diagnosis in under 150 seconds with an accuracy of 94.6\%. The results clearly show the potential of combining computational pathology and stimulated Raman histology for fast and accurate diagnostics in surgical procedures. 

Finally, due to its strategic positioning at the cross of molecular biology/omics, radiology/radiomics, and clinics, the rise of Computational Pathology -- by generating "pathomic" features -- is expected to play a crucial role in the revolution of spatial transcriptomics, defined as the ability to capture the positional context of transcriptional activity in intact tissue. Spatial transcriptomics is expected to generate a set of technologies allowing researchers to localize transcripts at tissue, cellular and sub-cellular levels by providing an unbiased map of RNA molecules in tissue sections. These techniques use microscopy and next-generation sequencing to allow scientists to measure gene expression in a specific tissue or cellular context, consistently paving the road toward more effective personalized medicine. Coupled with these new technologies for data acquisition, we have the release of new WSI brain datasets \cite{Roetzer-Pejrimovsky2022-sv}, new frameworks for deep learning analysis of WSI \cite{Berman2021-rb, Hagele2020-lt}, and methods to address the evergrowing concern of privacy and data sharing policies \cite{Lu2022-mc}.

\section*{Acknowledgments}

The human samples used in the preparation of the images throughout the chapter were obtained from the Neuro-CEB brain bank\footnote{Neuro-CEB brain bank : \url{https://www.neuroceb.org/en/}} (BRIF Number  0033-00011), partly funded by the  patients’  associations  ARSEP, ARSLA, “Connaître les Syndromes Cérébelleux”, France-DFT, France Parkinson and by Vaincre Alzheimer Fondation, to which we express our gratitude. We are also grateful to the patients and their families.

Knowledge and annotations have been provided with the support of Benoît Delatour and Lev Stimmer from the Alzheimer's Disease team - Paris Brain Institute (ICM), Paris, France.

Some of the research cited here (Section \ref{sec:understading}, and \ref{sec:methods}) was supported by Mr Jean-Paul Baudecroux, The Big Brain Theory Program - Paris Brain Institute (ICM) and the French government under management of Agence Nationale de la Recherche as part of the ``Investissements d'avenir'' ANR-10-IAIHU-06 (Agence Nationale de la Recherche-10-IA Institut Hospitalo-Universitaire-6).

Finally, the authors are grateful to María Gloria Bueno García for reviewing the chapter and providing useful comments.

\clearpage
\nocite{*}
\bibliographystyle{spbasicsort}
\bibliography{references}

\end{document}